\documentclass[twocolumn]{aastex63}

\usepackage{xspace}

\usepackage{mathtools}

\newcommand{\um}{$\rm\,\mu m$\xspace}

\submitjournal{\aj}
\accepted{October 13, 2021}

\shorttitle{Day-night temperature contrast of a strongly irradiated atmosphere}
\shortauthors{Lew et. al.}

\begin{document}

\title{Mapping the pressure-dependent day-night temperature contrast of a strongly irradiated atmosphere with HST spectroscopic phase curve}
\renewcommand*{\thefootnote}{\fnsymbol{footnote}}
\setcounter{footnote}{0}

  \correspondingauthor{Ben W. P. Lew }
  \email{lew@baeri.org}
\author[0000-0003-1487-6452]{Ben W. P. Lew}
  \affil{Lunar and Planetary Laboratory, The University of Arizona, 1640 E. University Blvd, Tucson, AZ 85721, USA}
\affil{Bay Area Environmental Research Institute and NASA Ames Research Center, Moffett Field, CA 94035, USA}
\author[0000-0003-3714-5855]{D\'aniel Apai}
  \affil{Lunar and Planetary Laboratory, The University of Arizona, 1640 E. University Blvd, Tucson, AZ 85721, USA}
  \affil{Department of Astronomy and Steward Observatory, The University of Arizona, 933 N. Cherry Ave., Tucson, AZ, 85721, USA}
\author[0000-0003-2969-6040]{Yifan Zhou$^{\ddag}$}\footnotetext{Harlan J. Smith McDonald Observatory Fellow}
    \affil{Department of Astronomy, University of Texas, Austin, TX 78712, USA}
\author[0000-0002-5251-2943]{Mark Marley}
  \affil{Lunar and Planetary Laboratory, The University of Arizona, 1640 E. University Blvd, Tucson, AZ 85721, USA}
  \affil{NASA Ames Research Center, Naval Air Station, Moffett Field, Mountain View, CA 94035, USA}
\author[0000-0002-4321-4581]{L. C. Mayorga}
\affil{The Johns Hopkins University Applied Physics Laboratory 11100 Johns Hopkins Rd Laurel, MD, 20723, USA}

\author[0000-0003-2278-6932 ]{Xianyu Tan}    
\affil{Atmospheric, Oceanic and Planetary Physics, Department of Physics, University of Oxford, OX1 3PU, UK.}

\author[0000-0001-9521-6258]{Vivien Parmentier}
\affil{Atmospheric, Oceanic and Planetary Physics, Department of Physics, University of Oxford, OX1 3PU, UK.}

\author[0000-0003-2478-0120]{Sarah Casewell$^{\S}$}\footnotetext{STFC Ernest Rutherford Fellow}
\affil{School of Physics and Astronomy, University of Leicester, University Road, Leicester, LE1 7RH, UK} 

\author[0000-0002-8808-4282]{Siyi Xu}
\affil{Gemini Observatory/NSF's NOIRLab, 670 N. A'ohoku Place, Hilo, Hawaii, 96720, USA}

\renewcommand*{\thefootnote}{\arabic{footnote}}

\setcounter{footnote}{0}

\begin{abstract}
Many brown dwarfs are on ultra-short period and tidally-locked orbits around white dwarf hosts. 
Because of these small orbital separations, the brown dwarfs are irradiated at levels similar to hot Jupiters.
Yet, they are easier to observe than hot Jupiters because white dwarfs are fainter than main-sequence stars at near-infrared wavelengths.
Irradiated brown dwarfs are, therefore, ideal hot Jupiter analogs for studying the atmospheric response under strong irradiation and fast rotation.
We present the 1.1--1.67\um spectroscopic phase curve of the irradiated brown dwarf (SDSS1411-B) in the SDSS J141126.20+200911.1 brown-dwarf white-dwarf binary with the near-infrared G141 grism of Hubble Space Telescope Wide Field Camera 3.
SDSS1411-B is a $50 M_{\rm Jup}$ brown dwarf with an irradiation temperature of 1300K and has an orbital period of 2.02864 hours.
Our best-fit model suggests a phase-curve amplitude of 1.4\% and places an upper limit of  11 degrees for the phase offset from the secondary eclipse.
After fitting the white-dwarf spectrum, we extract the phase-resolved brown-dwarf emission spectra.
We report a highly wavelength-dependent day-night spectral variation, with the water-band flux variation of about $360\pm70\% $ and a comparatively small J-band flux variation of $37\pm2\%$.
By combining the atmospheric modeling results and the day-night brightness-temperature variations, we derive a pressure-dependent temperature contrast.
We discuss the difference in the spectral features of SDSS1411-B and hot Jupiter WASP-43b, and the lower-than-predicted day-night temperature contrast of J4111-BD.
Our study provides the high-precision observational constraints on the atmospheric structures of an irradiated brown dwarf at different orbital phases.
\end{abstract}

\section{Introduction}

State-of-the-art phase curve observations of hot Jupiters \citep[e.g.,][]{knutson2012,cowan2012,arcangeli2019} find extreme temperature contrasts between dayside and nightside atmospheres that can exceed a few hundred Kelvin.
Such drastic day-night temperature contrasts in these strongly irradiated atmospheres drive a wealth of atmospheric waves, jets, and turbulence that affects the atmospheric composition and structure \citep[e.g.,][]{showman2002}.
The exciting observations of strongly irradiated atmospheres motivate a series of theoretical studies to understand the relevant physical and chemical atmospheric processes, including the radiative cooling and advection \citep[e.g.,][]{showman2002}, ohmic dissipation \citep[e.g.,][]{perna2010}, non-equilibrium chemistry \citep[e.g.,][]{agundez2014a},  non-homogeneous cloud formation and distribution \citep[e.g.,][]{parmentier2016,powell2018}, and hydrogen dissociation and recombination effects \citep{bell2018,komacek2018,tan2019} (see \citealt{heng2015,parmentier2018,showman2020,fortney2021} for reviews).
Despite the significant progress made on both the observational and modeling fronts on understanding the strongly irradiated atmospheres, many fundamental questions remained unanswered: What are the dominating physical mechanisms in redistributing irradiation energy at different rotation rates, temperatures, and altitudes? How do the clouds and atmospheric chemistry couple to global atmospheric circulation?

Time-series spectroscopy is one of the most powerful methods for constraining the atmospheric structure and tackling those fundamental questions.
By observing the emission at different orbital phases, we can study the hemispherically integrated spectra at different local times, or longitudes, of the irradiated atmosphere of a tidally locked planet.
The light curve profile of the spectroscopic phase curve constrains the day-night temperature contrast and possible atmospheric dynamics through comparisons to three-dimensional global circulation model results.
Furthermore, the wavelength dependence in the spectroscopic phase curve is a proxy of atmospheric properties at different altitudes because optical depth of an atmosphere is wavelength-dependent.
Therefore, spectroscopic phase curves have the potential to provide key insights into the variations of atmospheric structure at different longitudes and altitudes.

It is challenging, however, to obtain high-precision spectroscopic phase curves for hot Jupiters because of the bright host stars.
Almost all hot Jupiter phase curves are in photometric mode collected by Spitzer/IRAC with a few exceptions having been observed by the Hubble Space Telescope (HST) spectroscopic mode \citep[e.g.,][]{stevenson2014,arcangeli2019}.
Interestingly, the strong irradiation level experienced by hot Jupiters can also be found in a rare class of objects -- brown-dwarf white-dwarf systems with ultra-short orbital periods of only a few hours \citep[e.g.,][]{casewell2012,Beuermann2013,parsons2017,casewell2020a,casewell2020b}.
Because of the earth-sized white-dwarf radius and the white-dwarf spectrum, which often peaks at UV wavelengths, the near-infrared flux ratio of brown dwarf over white dwarf is higher than that of a hot Jupiter over its main sequence host star.
Therefore, the strongly irradiated brown dwarf in a brown-dwarf white-dwarf system is an easier target than hot Jupiters for obtaining a high-precision spectroscopic phase curve and is an excellent hot Jupiter analog.

In addition to being hot Jupiter analogs, studies of strongly irradiated brown dwarfs are important to understand the atmospheric dynamics in an atmosphere with similarly strong internal and external energy budgets \citep{showman2016}.
Advanced global atmospheric circulation simulations of brown dwarfs orbiting white dwarfs \citep{tan2020,lee2020} suggest that the rapid rotation of a few hours timescale causes a narrow equatorial jet, alternating eastward and westward zonal flows in off-equatorial regions, and large day-night temperature contrasts in the irradiated brown dwarf atmospheres.
The simulational results also indicate that the phase offset from the secondary eclipse\footnote{We follow the nomenclature of exoplanet community here even though the brown dwarf's emission is not fully obscured by the white dwarf at orbital phase of 0.5 (see also Figure \ref{fig:illustration}).} is pressure-dependent (see Figure 10 in \citealt{tan2020}). 
In addition to the atmospheric dynamic studies, the 1D atmospheric radiative transfer modeling of \citet{lothringer2020} finds that thermal inversion due to strong UV heating could explain the atomic emission lines seen in some of the hottest irradiated brown dwarf atmospheres \citep{longstaff2017}.
These models illustrate the rich and complex atmospheric dynamics and structure in the irradiated atmospheres.

To address the gap in our knowledge of the rapidly rotating irradiated atmosphere, we launched the observing campaign \textit{Dancing with the Dwarfs} (ID: 15947, P.I: D\'aniel Apai).
In this campaign, we use the HST Wide Field Camera 3 (WFC3)/G141 grism to obtain the spectroscopic phase curves of six irradiated brown dwarfs, which are SDSS-J141126.20+200911.1,GSC2.3-SBBO006229, SDSS-J155720.78+091624.7, BPS-CS-29504-0036, LSPM-J0135+1445, GD-1400, and SDSS-J155720.78+091624.7.
These irradiated brown dwarfs have equilibrium temperatures range from around 600 to 4000~K.
Here we present the results of the first object observed in this campaign, the irradiated brown dwarf in the SDSS J141126.20+200911.1 binary system.

\section{Brown-dwarf white-dwarf binary system SDSS J141126.20+200911.1}\label{sec:binary}
SDSS J141126.20+200911.1 (hereafter SDSS1411) is one of only three known eclipsing detached post-common envelope white dwarf-brown dwarf binaries. It was initially identified by \citet{Beuermann2013} in the Catalina Sky Survey and was further characterised by \citet{littlefair2014} and \citet{casewell2018}.  The white dwarf primary is a 13,000~K hydrogen-rich white dwarf with a mass of 0.53~M$_{\odot}$. The binary has likely been in this configuration for the 260$\pm$20 Myr cooling time of the white dwarf, although the space motion of the binary suggests it is a thin disk member with a total system age of at least 3~Gyr.  
The brown dwarf secondary SDSS1411-B is likely tidally locked \footnote{The tidal synchronization timescale is estimated be around 10Myrs based on Equation (1) in \citet{guillot1996} and an assumed tidal dissipation factor of $10^5$.} in its 121.73 min orbit and has a mass of 50~M$_{\rm Jup}$ and a radius of 0.70~R$_{\rm Jup}$ \citep{littlefair2014}.
The mass and radius of SDSSS1411-B are consistent with the brown dwarf evolutionary models \citep{baraffe2003} at the system age and indicates that the brown dwarf is not inflated. 
At an orbital distance of 0.003\,au, the irradiation temperature (assuming global redistribution of irradiation energy, zero albedo, and zero internal energy) of SDSS1411-B is about 1300\,K.
This brown dwarf mass and radius, suggests a spectral type of T5, while the lack of detected secondary eclipse and $K_{\rm s}$ band excess suggests a spectral type of L7-T1. \citet{casewell2018}
 used the infrared imager Hawk-I \citep{kissler-patig2008} of the Very Large Telescope to obtain the partial $J$-band and full $H$- and $K_s$-band phase curves of SDSS1411-B. They determined a day-night side temperature difference of $\sim$200~K based on the $H$-band phase curve. 
 By comparing the measured $H$ and $K_s$-band brightness temperatures to atmospheric models \citep{marley1999,marley2002,fortney2005}, they found a likely brightness temperature for the nightside of 1300~K, considerably warmer than would be expected by a T dwarf, particularly considering the high fraction of heat absorbed by the brown dwarf ($\sim$80 \% in the $K_{\rm s}$ band).

 The ground-based broadband photometric phase curves of SDSS1411-B provide important constraints on the phase curve amplitudes, yet they contain limited information on the possible dayside and nightside atmospheric structures.
 Resolved molecular features like water and methane absorptions from time-series spectroscopy is essential for constraining the temperature-pressure profile of the dayside and nightside atmospheres.
 By simultaneously probing different atmospheric depths with high-precision HST spectroscopic phase curve, our main goal is to answer the following question:
``How does heat redistribution vary across different pressures in the rapidly rotating irradiated atmosphere of SDSS1411-B?"

We structure the paper by first describing the data reduction of HST observational data in Section \ref{sec:reduction}. Based on the reduced time-series spectra, we analyze the light curve profile in Section \ref{sec:timeseries} and the spectral components in Section \ref{sec:spec}. We adopt a forward modeling approach to model the irradiated atmosphere and interpret the observed spectra in Section \ref{sec:models}. Finally, we discuss the implications of our results and compare the spectra of the irradiated brown dwarf with that of hot Jupiters and isolated brown dwarfs in Section \ref{sec:discussion}.

\section{Data Reduction}\label{sec:reduction}
We observed SDSS1411 in six consecutive Hubble Space Telescope (HST) orbits with WFC3/IR/G141 grism on Jan 4th 2020 in the HST program GO-15947 (PI: Apai). In each HST orbit we obtained two to three F127M direct images for wavelength calibration and eight G141 Multiaccum exposures.
Each Multiaccum exposure spans 313\,s and consists of fifteen read-outs. 
Given the short orbital period of the target system, we aimed to retain the highest possible time resolution in our analysis. Therefore, we performed the data reduction starting from the \textit{\_ima} files, which store every single detector read-out.
The \textit{$\_$ima} files are products of the standard \textit{calwfc3} pipeline, which removes dark current,  corrects the bias, subtracts zero-read signals, calibrates the detector non-linearity, flags bad pixels, and estimates electron count uncertainty. 
In each \textit{\_ima} file, the recorded count rates and uncertainties are cumulative averages with the zeroth readout as the reference point.
We calculated the difference in counts and the corresponding uncertainty between two adjacent read-outs, obtaining --  in each exposure -- fourteen $22.35\rm\,s$ independent differential frames. In total there are 672 frames over the six HST orbits.

We masked all pixels flagged for bad data quality (DQ), except those that were only flagged with cosmic-ray hits. These pixels were instead re-processed via our own cosmic-ray detection algorithm.
In our algorithm, we identified cosmic-ray impacted pixels based on the change in count rate, which is measured by the difference in electron counts between two successive frames.
We find that a threshold of five times of the photometric uncertainty can separate most of the drastic changes in count rates due to cosmic-ray events from those that are due to the white dwarf-brown dwarf eclipse event (the fastest change in the system's intensity due to a physical process, i.e., $\sim 2$ minutes). We verified the results by visually inspecting the maximum electron count and the number of detected cosmic-ray events of each pixel in and out of the eclipse event.

We removed the sky component by least chi-squares fitting a scaled master sky image \citep{kummel2011} and subtracted it from the science images.
Based on the fitted centroid positions of SDSS1411 in our direct images, we calculated and applied the field-dependent flat field correction. After the sky removal and flat-fielding, we interpolated the masked pixels from unmasked pixels within the same row.
We then corrected the HST ramp effect by using the RECTE ramp-correction method as described \citet{lew2020b}, which solves the intrinsic incoming count rate by minimizing the difference in count rates between the RECTE model prediction \citep[][]{zhou2017} and the observations.

\subsection{Spectral Extraction}
Because the standard spectral extraction pipeline \textit{aXe} \citep{kummel2009} does not directly work with the \textit{\_ima.fits} files, we developed our own spectral extraction procedure that follows the method described in the \textit{aXe} manual.
In addition, we included the optimal spectral extraction algorithm \citep{horne1986} in our pipeline, as described below.

First, based on the SDSS1411 position on the direct images and the position-dependent wavelength solution by \citet{kuntschner2009}, we constructed a wavelength solution for the spectral traces in the G141 images.
Afterwards, we converted counts-per-pixel units into counts-per-wavelength units by considering a weighting function, which calculates the fractional area of pixels projected to wavelength solutions in the cross-dispersion direction (see Appendix \ref{sec:weighting}).
We also used the median-averaged spectra over six HST orbits as the two-dimensional spectral dispersion profile to extract the spectra with the optimal extraction method \citep{horne1986}.
We use an aperture size of eight pixels wide in the cross-dispersion direction for the spectral extraction.
We converted the electron counts to flux densities by dividing them by the sensitivity curve\footnote{\url{https://www.stsci.edu/hst/instrumentation/wfc3/documentation/grism-resources/wfc3-g141-calibrations}}, considering the wavelength resolution and the exposure time.
Finally,  to correct the wavelength-dependent flux loss due to the finite aperture of eight pixels, we performed an aperture correction using an interpolated table based on \citet{kuntschner2011}.

\section{Light Curve Analysis}\label{sec:timeseries}
\begin{figure*}[!htp]
    \centering
    \includegraphics[width=1.05\textwidth]{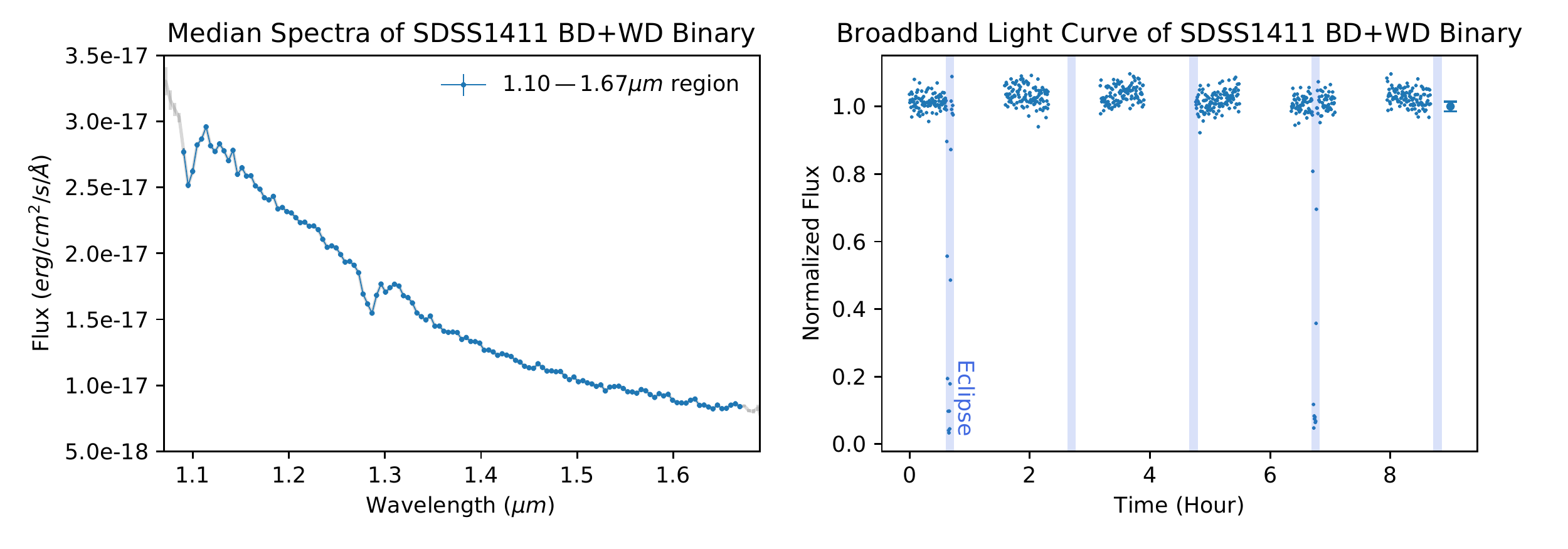}
    \caption{\textit{Left panel:} The median spectra of the white dwarf-brown dwarf system. The blue line highlights the spectra from 1.1 to 1.67\um. The two hydrogen absorption regions of white dwarf are $1.09$--$1.11$\um and $1.27$--$1.32$\um.  \textit{Right panel:} The six-orbit $1.1$--$1.67$\um integrated broadband light curve. The two white dwarf eclipse events occur at the first and the fifth HST orbits. The error bar at $t=9$~hours shows the representative flux uncertainties of 1.4\% for single photometric point. The shaded blue regions indicate the expected eclipse events with a period of 121.72 minutes.}
    \label{fig:overview}
\end{figure*}

Figure \ref{fig:overview} provides an overview of the reduced white dwarf and brown dwarf combined spectra and light curve. The spectra show the Paschen-beta and Paschen-gamma hydrogen absorption features of the white dwarf spectrum at around $1.095${\um} and $1.284$\um. We detect two primary eclipse (occultation of white-dwarf emission by brown dwarf) events in the six HST-orbit observation. 

\subsection{Day-night contrast and mid-eclipse time} \label{sec:mideclipse}
\begin{figure*}[!h]
    \centering
    \includegraphics[width=1.0\textwidth]{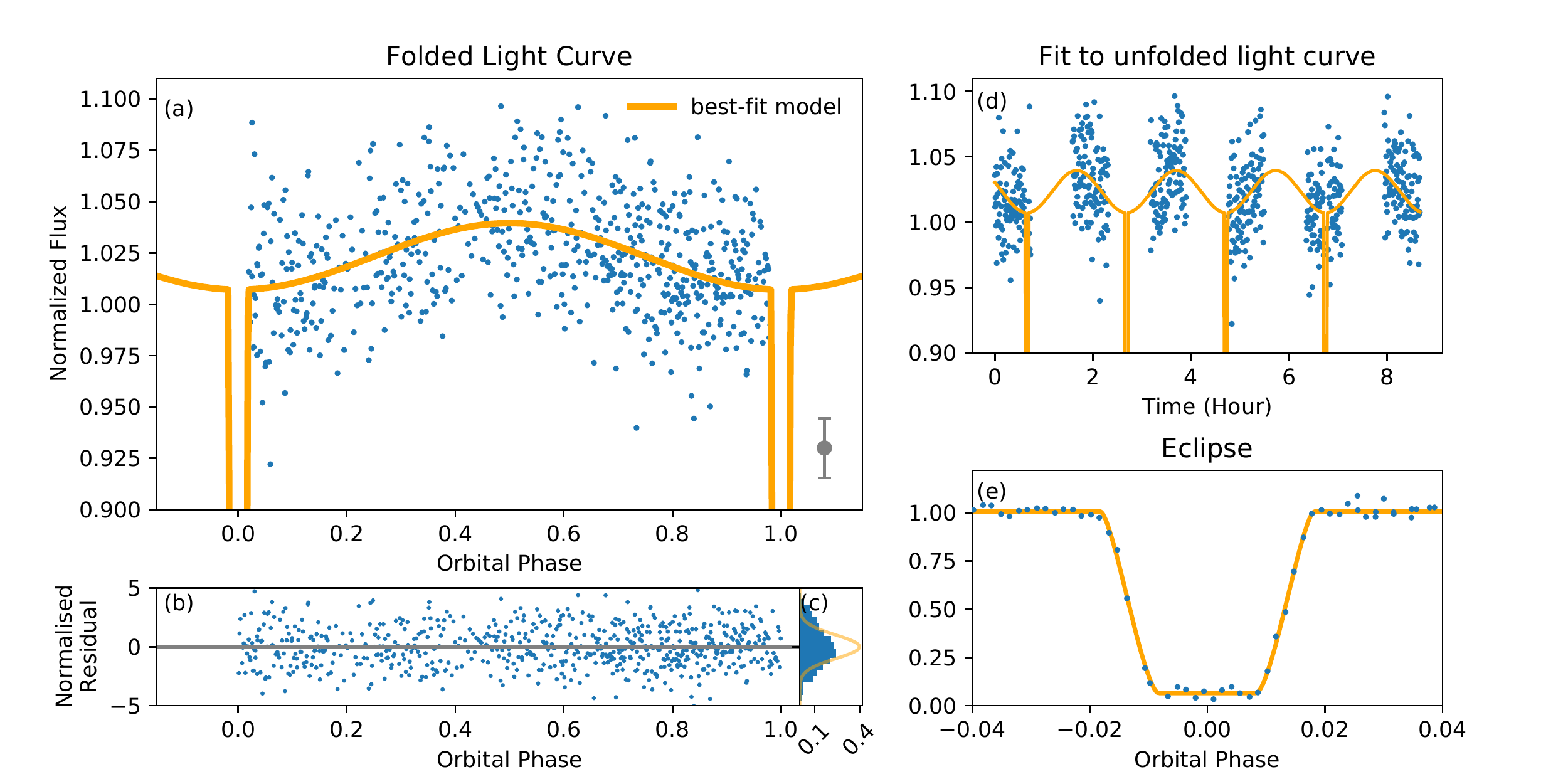}
    \caption{\textit{(a):} The folded light curve with the fitted period of 2.02862 hours. The primary eclipse part of the phase-folded light curve is plotted separately in the panel (e).  \textit{(b):} The residuals of the model fit to the light curve. The residuals are normalized by the photometric uncertainty. \textit{(c):} The histogram of all the residuals. The orange line shows a normal distribution with one standard deviation. \textit{(d):} The best-fit model and the light curve plotted in units of time on the x-axis.  \textit{(e):} A zoom-in plot of the model fit to the light curve in the eclipse region.}
    \label{fig:sinefit}
\end{figure*}

\begin{figure}[!h]
    \centering
    \includegraphics[width=0.5\textwidth]{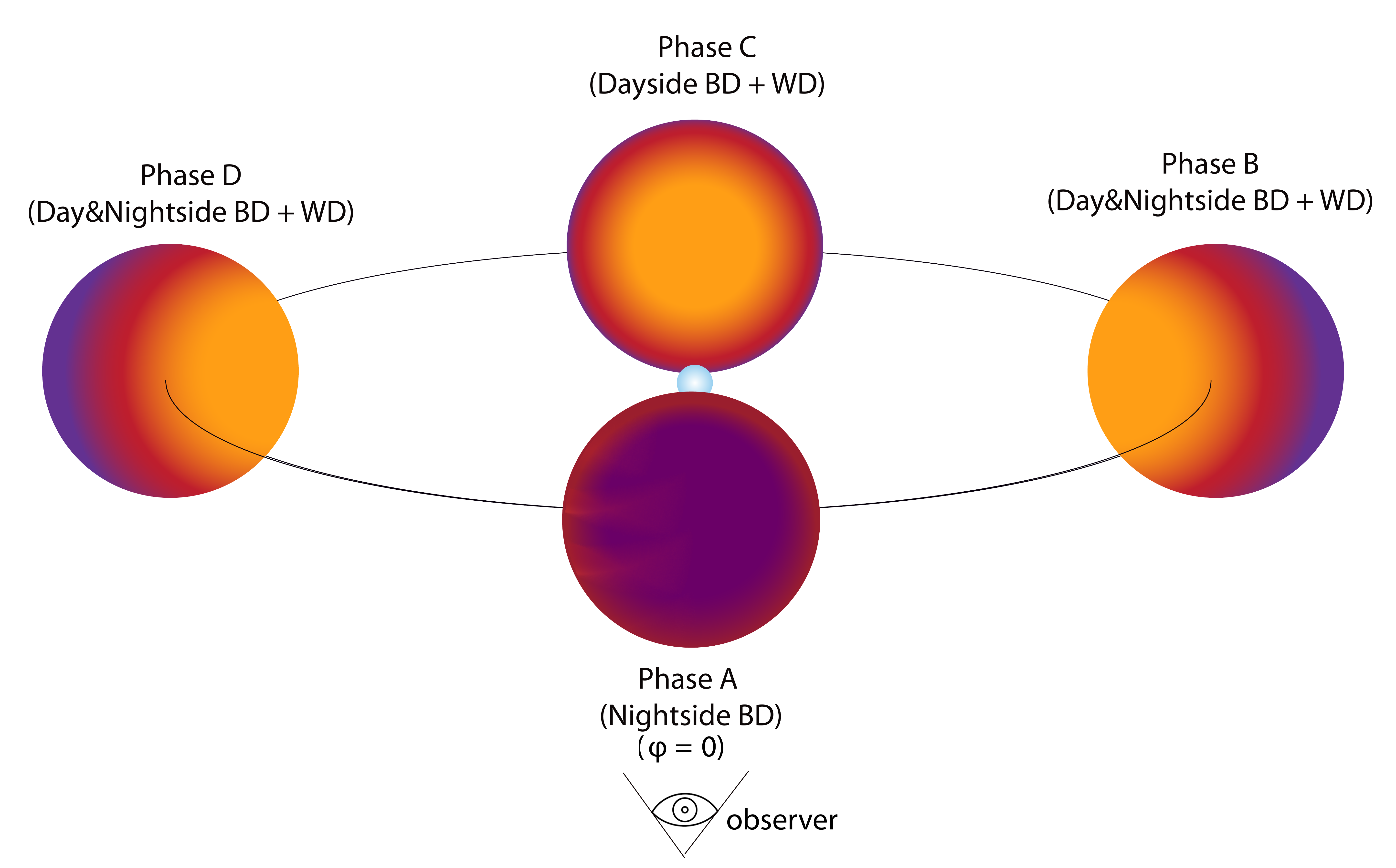}
    \caption{An illustration of the white-dwarf brown-dwarf system at different phases. We define Phase A ($\phi =0$) as the primary eclipse phase and Phase C ($\phi = 0.5$) as the secondary eclipse phase. The white-dwarf emission is completely blocked by brown dwarf, so only the brown dwarf nightside (midnight) emission is observed during the primary eclipse. Because the brown dwarf radius is at least 3.5 times larger than that of the white dwarf, the brown dwarf's emission is at best only partially obscured during the secondary eclipse.}
    \label{fig:illustration}
\end{figure}

We integrate the spectra from $1.15$--$1.67$\um to create a synthetic broadband white light curve. We normalize the white light curve with the mean of light curve.
We construct a simple model to fit the normalized light curve:
\begin{align*}
    \text{Model(t)} = M_{\text{eclipse}}(t,P, i,t_0,r_p/r_s, a, u) \times c_{\mathrm{WD}} +  \\ A \sin[\frac{2\pi (t-t_0)}{P} + \phi_{\mathrm{sine}}] + c_{night} \text{,}
\end{align*}
 where $M_{\text{eclipse}}$ represents the eclipse model calculated with \texttt{Batman} package \citep{kreidberg2015}.
 There are six parameters in $M_{\text{eclipse}}$ -- orbital period $P$, mid-eclipse time $t_0$, inclination $i$, the ratio of the brown dwarf radius over the white dwarf radius $r_p/r_s$, the orbital semi-major axis $a$, a linear limb darkening coefficient $u$.
 The modeled white-dwarf flux contribution equals to $M_{\text{eclipse}}$ multiplied by the time-averaged  broadband white-dwarf flux $c_{wd}$.
 The sinusoidal term and the constant $c_{\mathrm{sine}}$ term represents the orbital-phase modulated brown dwarf's emission. $\phi_{\mathrm{sine}}$ is defined as the phase offset of the primary-eclipse time from the zero point of the sine wave.

We plot the best-fit light-curve model with least chi-square method in Figure \ref{fig:sinefit} with fitted parameters in Table \ref{table:eclipse}.
The fitted reduced chi-square is 3.1. 
The histogram of the residuals normalized by photometric uncertainties in panel (c) has a standard deviation of 1.76.
Compared to spectra extracted with uniform aperture, the optimal extraction method lowers the averaged photometric uncertainty from 4\% to 1.8\% and gives a similar median (-0.001) and standard deviation (0.025) of the residuals of light-curve fitting. 
We cannot reject the null hypothesis (i.e., normal distribution) based on the Shapiro-Wilk normality test of the residual with a p-value of 0.87.
Therefore, we argue that the sinusoid model is a good approximation to the light curve out of eclipse.

We use the Markov chain Monte Carlo (MCMC) method with \texttt{emcee} \citep{foreman-mackey2013} to sample the posterior distributions of the fitted model parameters.
We set up the MCMC chain with 400 walkers for 400,000 steps. 
We listed the uniform priors of parameters in Appendix \ref{sec:mcmc}. 
We find that the fitted radius ratio, semi-major axis, and inclination are correlated; the fitted eclipse baseline and sinusoidal baseline are correlated, too.
The auto-correlation timescales of all parameters are less than 0.3\% of the total steps, and thus are well converged.
We listed the 15, 50, and 85 percentiles of the posterior distributions of the model parameters in Table \ref{table:eclipse}.

Based on the marginalized posterior distribution of $c_{\mathrm{WD}}$ and $c_{\mathrm{sine}}$, the mean broadband flux of the white dwarf is about 0.94/0.08 $\sim$ 12 times higher than the brown dwarf's time-averaged emission.
If we assume that the amplitude of light curve solely originates from the brown dwarf's emission variation, then the intrinsic brown dwarf variability amplitude is about 19\%, equivalent to a 38\% peak-to-trough flux variation.
Assuming the photosphere radius of the dayside and nightside of SDSS1411-B atmosphere are identical \footnote{The light curve fitting results of the g’-band primary-eclipse in Littlefair et al. (2014) find no detection of reflection from SDSS1411-B and see no evidence of Roche-lobe distortion. We used the ROCHE software, which is part of the LCURVE model \citep{copperwheat2010}, to estimate the Roche distortion due to tidal force and rotation. Based on the given mass ratio (0.1), orbital separation (0.003au), and brown dwarf radius (0.7 $R_{\mathrm{Jup}}$), the modeled distortion for SDSS1411-B is around 0.67\%. The model calculation is the same as that in \citet{casewell2020a}. 
}, then the {38\%} broadband flux variation corresponds to a day-night temperature variation of about 8\%.

Based on the marginalized posterior distribution of the phase of sinusoid, the phase of mid-eclipse is ($1.50 \pm 0.02)\, \pi$. 
The fitted sinusoidal phase suggests that the mid-eclipse occurs at the trough of the fitted sinusoid.
The fitted orbital period of $2.02863\pm 0.00008$ hours, or 121.718 $\pm$ 0.005 minutes is within 2-$\sigma$ of the reported period of 121.73 minutes in \citet{Beuermann2013}.

\begin{table}[]
\raggedright
\begin{tabular}{lll}
\hline
\hline
& least $\chi^2$ & MCMC \\ \hline
Orbital period P (hour) & 2.02864 & 2.02863  $\pm 0.00006$      \\
Radius ratio $r_{\text{BD}}/r_{\text{WD}}$  & 4.6& $3.6^{+1.5}_{-0.8}$            \\ 
Inclination $i$ (deg)      & 86 & $87^{+2}_{-2}$  \\ 
Mid-eclipse time $t_0$ (hour)& 0.65388 & $0.6539 \pm 0.0001$      \\ 
Semi-major-axis ratio a/$r_{WD}$ & 43&  $37^{+7}_{-4}$         \\ 
Limb-darkening coefficient &  0.0& $0.3\pm0.2$  \\
Flux constant $c_{WD}$ & 0.943 &       $0.943 \pm 0.004 $        \\ 
$A_{\mathrm{sine}}$    &0.0162   &   $0.0160 \pm 0.0008$        \\ 
$c_{\mathrm{sine}}$ &0.081 & $0.080 \pm 0.004$\\
$\phi_{\mathrm{sine}} $ & 1.50$\pi$    & ($1.50 \pm 0.02$)$\pi$             \\ 
\hline \hline
\end{tabular}
\caption{The second column shows the best-fit model parameters from a least chi-squares method. The third column shows the median values and 1$\sigma$ uncertainties (16 and 84 percentiles)  of the marginalized posterior distributions of model parameters with the MCMC method. The semi-major axis ratio, inclination, and radius ratio are correlated. }\label{table:eclipse}
\end{table}

\subsection{Phase offset from the secondary eclipse phase} \label{sec:phaseoffset}
\begin{figure}[!h]
    \centering
    \includegraphics[width=0.5\textwidth]{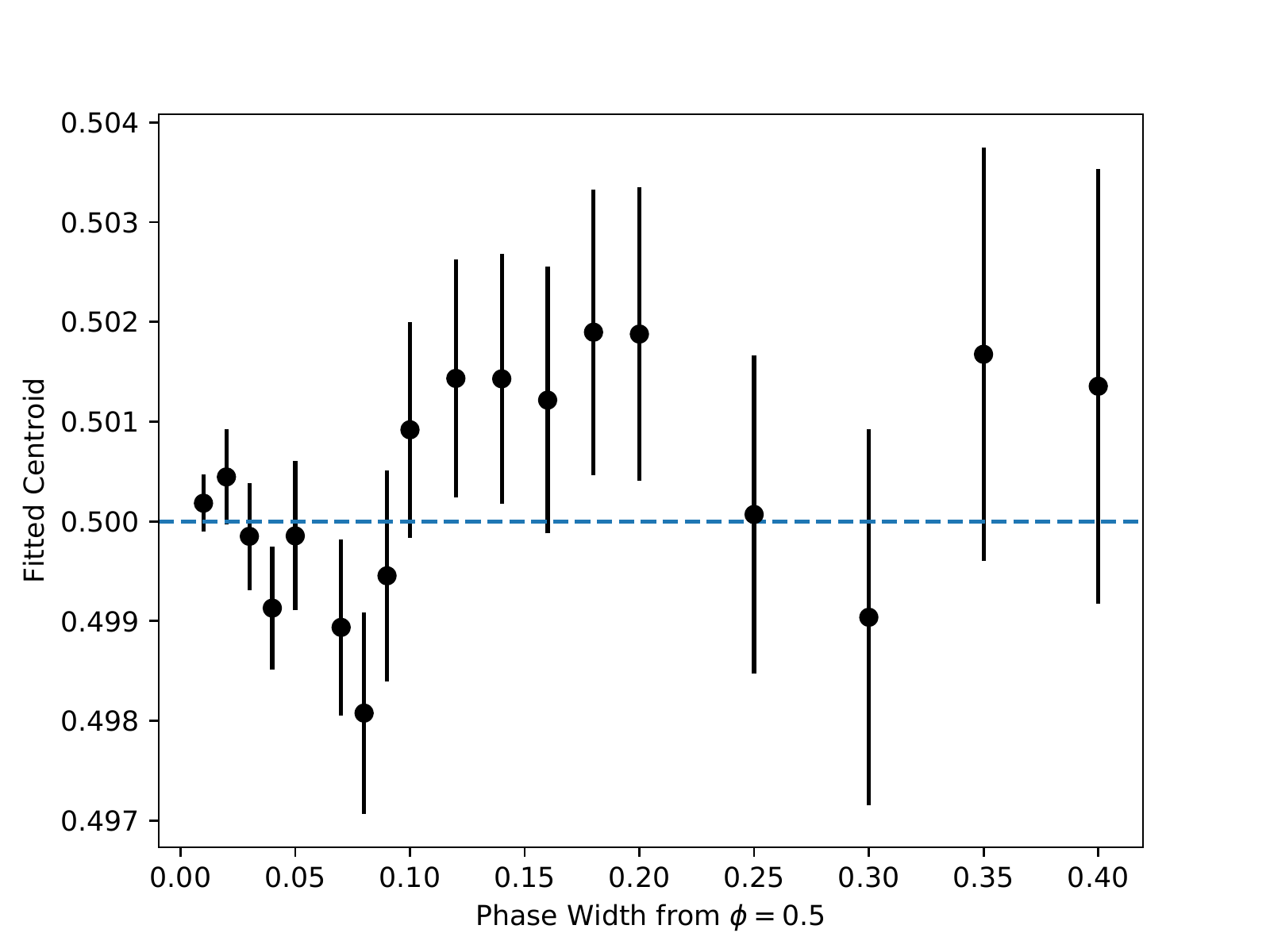}
    \caption{The calculated centroids with partial light curves with phase coverage around the secondary eclipse ($\phi=0.5$). The centroid of a light curve that peaks after secondary eclipse has a value above 0.5 and vice versa. The derived centroids depend on the phase coverage of the partial light curves, so we find no convincing evidence of phase offset based on our dataset.}
    \label{fig:centroid}
\end{figure}

The fitted phase of the mid-eclipse time suggests no significant phase shift between the mid-eclipse time and the trough of fitted sinusoid.
By fitting the full-orbital-phase light curve, our light curve model is less sensitive to the potential phase offset between the expected secondary eclipse phase (orbital phase of 0.5) and the peak of light curve.
Therefore, we use a non-parametric method to examine if there is any phase offset from the secondary eclipse.
We calculate the centroids of partial light curves with different phase coverages.
We estimate the centroid uncertainty by adding photometric noise to the partial light curve and repeat the calculation for 10,000 times.
We use the 16 and 85 percentiles of the 10,000 centroid samples as our centroid uncertainties for the partial light curves.
In Figure \ref{fig:centroid}, our results suggest that the calculated centroids depend on the phase coverages of the partial light curves.
The fitted centroids are well within the uncertainty of the fitted eclipse phase ($\sigma_{\phi_{\mathrm{sine}}} = 0.02\pi$ = 0.01 orbital phase, see Table \ref{table:eclipse}).
Therefore, we conclude that we find no evidence of phase offset from the secondary eclipse that exceeds a 3-$\sigma$ limit of 0.03 orbital phase.
The nearly zero phase offset suggests that the observed hemisphere appears to be the hottest at the secondary eclipse, or local noon time.
This suggests that the radiative cooling timescale of the near-infrared photosphere is shorter than the horizontal advection timescale. Alternatively, the spatial scale of the circulation patterns could be too small to affect the phase offset.

We also fit the single sinusoidal model to the $J'$-band, water-band, and the $H'$-band light curves. We conclude that no significant phase offsets (less than 3\% of orbital phase) are detected between the three light curves and the white light curve.

\subsection{Brown dwarf radius}\label{sec:bdradius}
\begin{figure}[!h]
    \centering
    \includegraphics[width=0.5\textwidth]{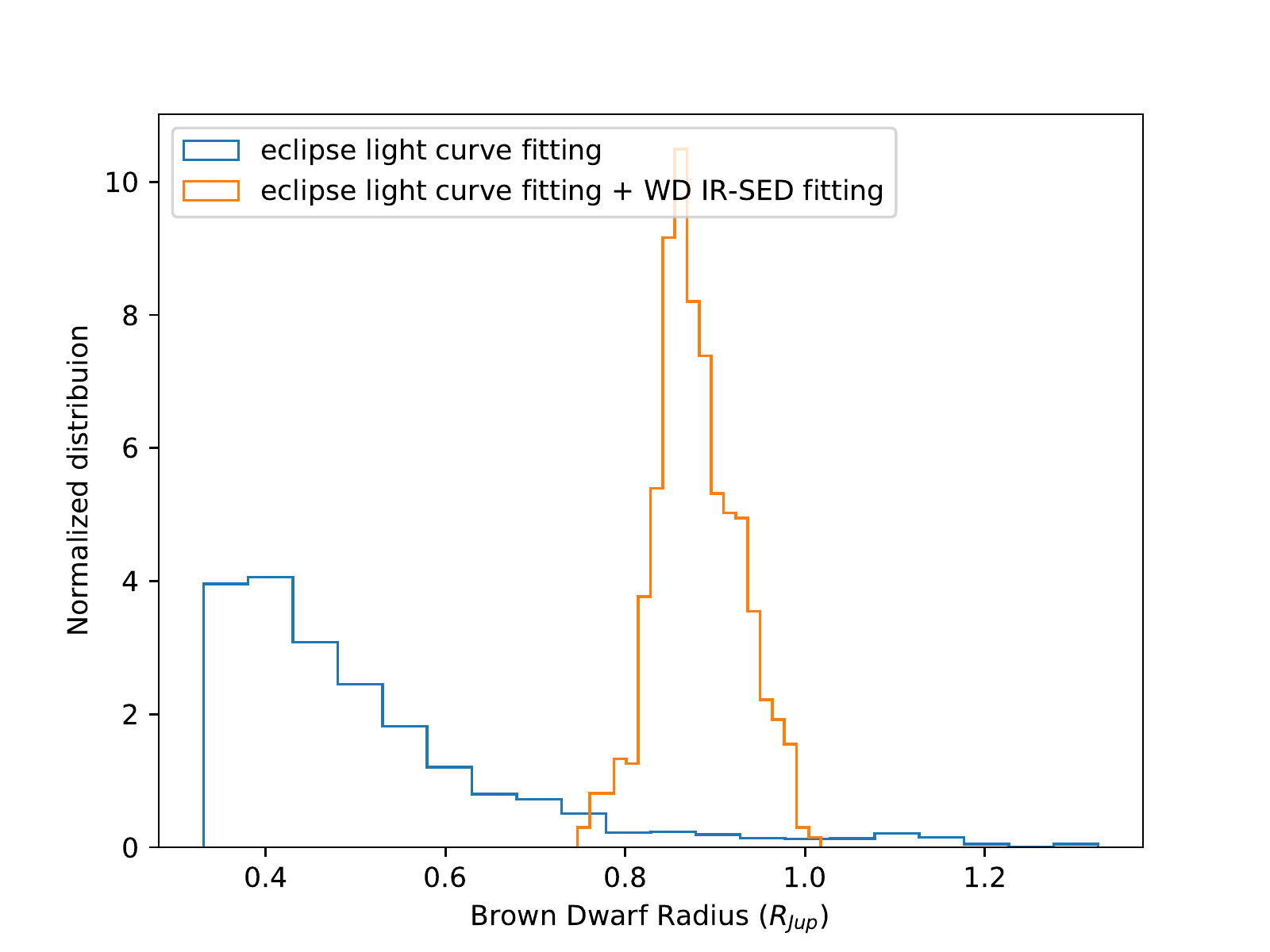}
    \caption{The comparison of marginalized brown-dwarf radius posterior distribution based on the MCMC eclipse light curve fitting results and that with the additional constraint from white-dwarf evolutionary models, radial velocity semi-amplitudes, and white-dwarf near-infrared spectral fitting.}
    \label{fig:bdradius}
\end{figure}

Our eclipse light curve alone does not uniquely constrain the brown dwarf radius because of the strong degeneracy between inclination, white-dwarf brown-dwarf radius ratio, and semi-major axis.
We utilize multiple independent measurement methods to further constrain the brown dwarf radius.
The absolute white-dwarf radius is $0.0130 \pm 0.0003 R_{\odot}$ based on near-infrared spectral fitting from Section \ref{sec:night}.
Given the absolute white-dwarf radius and the estimated SDSS1411 system mass of $0.58 \pm 0.03 \, M_{\odot} $  that is derived from the white-dwarf evolutionary models and the semi-amplitude of radial velocity measurements in \citet{littlefair2014}, we constrain the semi-major axis over white-dwarf radius ratio to be $a/r_{WD} = 52 \pm 2$.
By sampling the posterior distribution of light-curve fitting that satisfies the semi-major axis constraints, we plot the updated marginalized posterior distribution of brown dwarf radius in Figure \ref{fig:bdradius}.
The 16, 50, and 84 percentiles of the brown dwarf radius distribution are 0.84, 0.87, and 0.93 $\, \textrm R_{\mathrm{Jup}}$ respectively.
Our derived brown dwarf radius values are higher than that in \citet{littlefair2014} by $\sim 2\sigma$ mainly because we adopt a closer distance (177 vs. 190 pc).
Since the common envelope evolution history and the tidal interaction in the SDSS1411 system, it is possible that the radius and the mass of SDSS1411-B could be different than those predicted by evolutionary models at a given age.
We find that the updated brown dwarf radius  of 0.87$\pm 0.02 R_{\rm Jup}$ is consistent with the 0.854--0.807 $R_{\rm Jup}$ at ages of 3--10 Gyr respectively predicted by brown dwarf evolutionary models Sonora V2.0 (Marley et al., submitted) for a 50 $M_{\rm Jup}$.

\section{Spectral Analysis}\label{sec:spec}
Before we can study the dayside and nightside spectra of the irradiated brown dwarf atmosphere, we need to separate out the white-dwarf spectrum from the spatially unresolved white-dwarf and brown-dwarf combined spectra.
We assume that there are two stationary and one time-dependent components in the spectroscopic phase curves: the white dwarf spectrum $F_{\text{WD}}$, the brown dwarf dayside spectrum $F_{\text{dayBD}}$, the brown dwarf nightside spectrum $F_{\text{nightBD}}$, and the dayside area fraction $A(\phi)$ as a function of orbital phase $\phi$:
{\setlength{\belowdisplayskip}{-10pt}%
\setlength{\abovedisplayskip}{10pt}%
\begin{align*}
    F_{\text{total}}  = F_{\text{WD}} + F_{\text{nightBD}} + [1-A(\phi)] \times (F_{\text{dayBD}} - F_{\text{nightBD}})
    \shortintertext{where $\phi$ = [0,1)} 
\end{align*}
}
In Section \ref{sec:timeseries} we already solved for the wavelength-averaged  $A(\phi)$.
In the following subsections, we derive each spectral component based on the model.

\subsection{Extracting white dwarf spectrum and brown dwarf nightside emission}\label{sec:night}
      \begin{figure}[!h]
        \centering
        \includegraphics[width=0.5\textwidth]{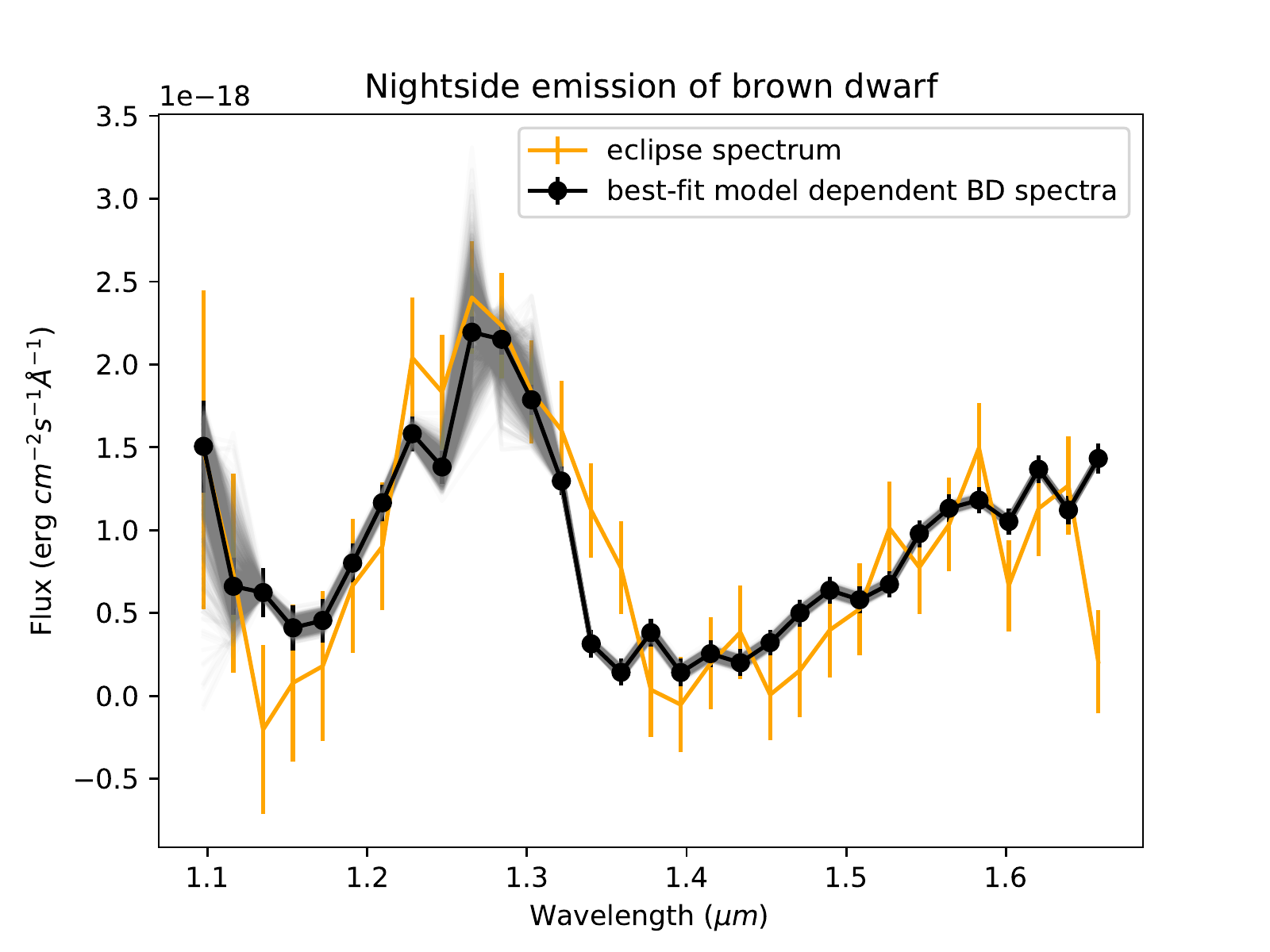}
        \caption{The binned averaged nightside spectrum (black line)observed just before or after the primary eclipse, respectively after the removal of white-dwarf spectrum shows less scatters between spectral points and is similar to the SDSS1411-B spectrum during eclipse (orange line). The grey lines show the potential bias of the averaged nightside spectrum due to uncertainty of white-dwarf temperature, gravity, and wavelength calibration sampled by the Monte Carlo method.}\label{fig:eclipse2}
    \end{figure}
When the white dwarf is eclipsed by the brown dwarf at $\phi = 0$, the observed flux ($F_{\text{eclipseBD}}$) is equal to the brown dwarf's nightside emission. 
In Figure \ref{fig:eclipse2}, we plot the brown dwarf's emission during the eclipse.  
We report 11-$\sigma$ and 7-$\sigma$ level detections in the 1.2--1.3\um and 1.5--1.6\um regions, respectively.
  
We obtained a nightside spectrum with a higher signal-to-noise ratio than that of the eclipse spectrum by leveraging our understanding of the white dwarf spectrum. 
We apply a model-driven approach to separate the nightside emission from the white dwarf spectrum during the night phase, which is defined as the phase intervals before and after the eclipse event ($\phi =$ 0.90--0.98 and 0.02--0.07).
During the night phase, the observed spectra are assumed to be
\vspace{-1cm}
\begin{align}
    \shortintertext{$\phi_{\text{nightBD}} = [0.9$--$0.98, 0.02$--$0.10];$\vspace{-.95cm}} \nonumber \\ \shortintertext{$\phi_{\text{eclipseBD}} = [0.992$--$1, 0$--$0.008] $\vspace{-.5cm}} \nonumber \\
    F_{\text{nightphase}}  &= F_{\text{WD}} + F_{\text{nightBD}} \\
    \shortintertext{If $F_{\text{nightBD}} \approx F_{\text{eclipseBD}}$,\vspace{-.5cm}} \nonumber \\
    F_{\text{WD}} &\approx F_{\text{nightphase}} - F_{\text{eclipseBD}} \label{eq:nightwd}
\end{align}

\begin{figure}[!h]
    \centering
    \includegraphics[width=0.5\textwidth]{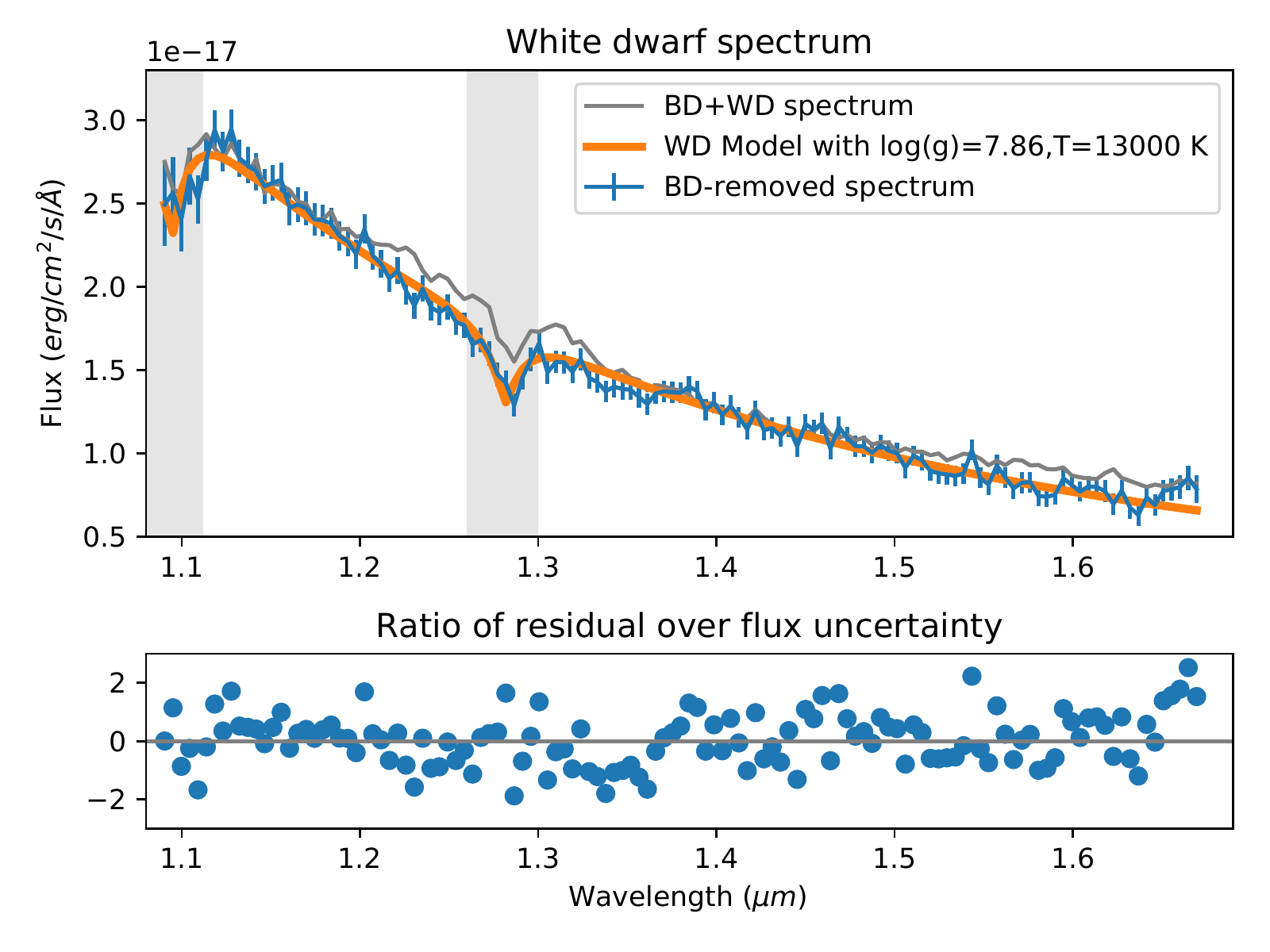}
    \caption{The blue line shows the cleaned white dwarf spectrum after removal of brown dwarf eclipse spectrum. The orange line shows the binned white-dwarf model spectrum. The grey shaded regions highlight the absorption lines in the white dwarf spectrum.}
    \label{fig:whitedwarf}
\end{figure}

In this approach, we assume that the brown dwarf's emission remains unchanged throughout the night phase and approximate the brown dwarf's spectrum at night phase with the eclipse spectrum.
We also assume a constant (phase-independent) white dwarf emission.
Therefore, we can extract the white-dwarf spectrum by subtracting the approximated brown dwarf's night-phase  spectrum from the combined spectrum (see Equation \ref{eq:nightwd}).

We adopt the best-fit white-dwarf model parameters (log(g) = 7.86$\pm 0.07$, T=13000$\pm 300\,K$) reported in \citet{littlefair2014}.
We linearly interpolate the white-dwarf model spectra grid of \citep{koester2010} to obtain the white-dwarf model spectrum with log(g) = 7.86 and T=13000\,K.
We then obtain the best-fit white-dwarf model spectrum by fitting the white-dwarf model spectrum to the white-dwarf spectrum approximated from Equation \ref{eq:nightwd}.
The only free variable in the spectral fitting is the scaling factor, $r_{\text{WD}}^2/d^2$, where $r_{\text{WD}}$ is the white-dwarf radius and $d$ is the distance of the SDSS1411 system.
We plot the white-dwarf model fitting results in Figure \ref{fig:whitedwarf}.
We note that the center of Paschen-beta and Paschen-gamma absorptions in the observed spectra is off by one pixel (0.0046\um) than that of the model spectra, which is likely caused by imperfect absolute wavelength calibration.
Our subsequent analysis focuses on the broadband features so the slight deviation does not affect our results.
At a GAIA DR2 distance of $177\pm 5\,$pc, the fitted scaling factor indicates a white-dwarf radius of $0.0130 \pm 0.0003\, \textrm R_{\odot}$.

Based on the best-fit white-dwarf model spectrum, we then derive the brown dwarf's nightside spectrum as shown by the following equation:
\begin{align}
\centering
    F_{\text{nightBD}} = F_{\text{nightphase}} - 
    (r/d)^2 F_{\text{modelWD}}\label{eq:nightbd}
\end{align}
Because the phase coverage of spectra observed during the night-phase ($\Delta \phi = 0.16 $) is wider than that during the eclipse phase ($\Delta \phi = 0.016$), the averaged night-phase spectra has a much higher signal-to-noise ratio than the eclipse spectrum, as shown in Figure \ref{fig:eclipse2}.

The accuracy of the model-dependent night-phase spectrum depends on the accuracy of white-dwarf temperature and gravity and that of the wavelength calibration of the HST data.
To investigate the potential systematic uncertainties in our results, we use a Monte-Carlo method to sample three one-dimensional normal distributions of temperature, gravity, and wavelength offset.
We adopt the uncertainties ($\sigma_T=300K$, $\sigma_{\log(g)} = 0.07$) reported in \citet{littlefair2014} as the standard deviations of the normal distributions of temperature and gravity.
We conservatively estimate the uncertainty in the wavelength calibration to be half the pixel resolution, or 23$\rm \AA$.
Based on the 1000 white-dwarf model spectra that derived from the three normal distributions of temperature, gravity and potential wavelength offset, we fit each model to the observed white-dwarf spectrum (Eq. \ref{eq:nightwd}) with the least chi-squared method.
The scaling factor is the only free parameter in the fitting process.
In Figure \ref{fig:eclipse2}, we plot the derived brown-dwarf night-phase spectral component after subtracting off the fitted white-dwarf model spectra from the observed night-phase spectra.
We show that, except at wavelengths where white-dwarf model spectra show strong hydrogen absorption lines (1.25--1.30\um and 1.09--1.11$\,\mu m$), the model-dependent night-phase spectrum of brown dwarf is robust against the white-dwarf temperature, white-dwarf gravity, and the HST wavelength calibration uncertainties.

\subsection{Phase-resolved spectra and day-night spectral variation}\label{sec:specvar}

\begin{figure}
    \centering
    \includegraphics[width=.46\textwidth]{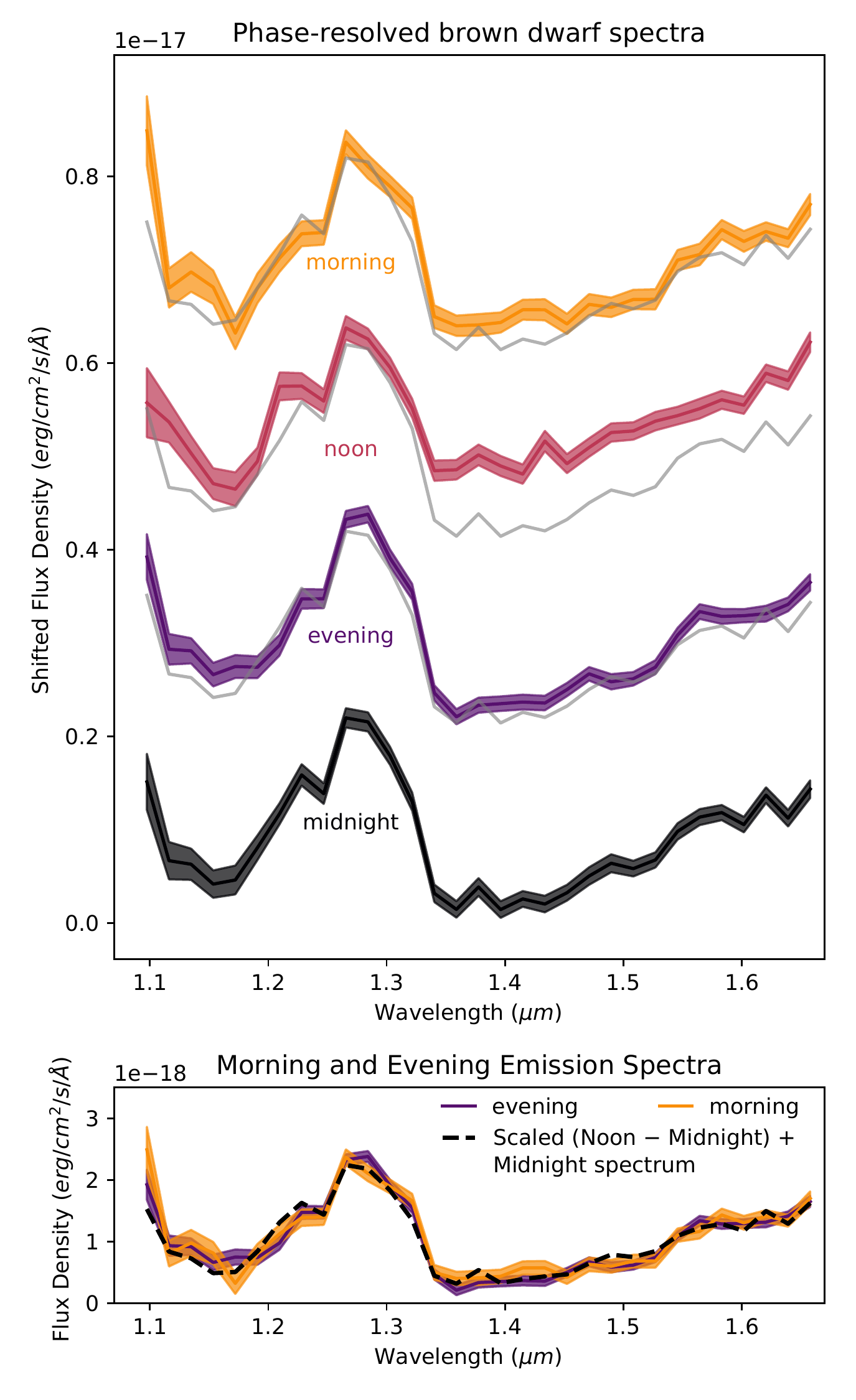}
    \caption{Top panel: the binned spectra at morning, noon, evening, and night orbital phases. The grey line represents the nightside spectrum. Bottom panel: Comparison of the evening and morning spectra shows similar spectral features across 1.1--1.67\um. The scaled day-night spectral difference (dahsed line) also matches well to the evening and morning spectra, showing that the evening and morning spectra can be reproduced by a linear combination of the dayside and nightside spectra.}
    \label{fig:phasespec}
\end{figure}

After fitting and subtracting the phase-independent white-dwarf spectral component, we plot the derived phase-averaged brown dwarf spectra over the midnight ($\phi$ = 0.90--0.98 and 0.02--0.10), evening ($\phi$ = 0.17--0.33), noon ($\phi$ = 0.42--0.58), and morning ($\phi$ = 0.67--0.83) orbital phases in Figure \ref{fig:phasespec}.
We test if the evening and morning spectra is a linear combination of dayside (noon) and nightside (midnight) spectral component.
We fit a scale factor $k$ to the composite spectra made of dayside and nightside spectra to the observed evening and morning spectra as shown by the cost function equation below:
\begin{equation}
    C = [k \times (F_{\rm day} - F_{\rm night}) + F_{\rm night} ] - \\ F_{\rm  evening} ({\rm or }\,   F_{\rm morning})
\end{equation}
Our least-square fitting that minimizes the cost function $C$ over observational uncertainties give reduced chi-squares of 0.98 and 0.88.
The close-to-unity reduced chi-squares suggest that both evening and morning spectra can be well approximated by a linear combination of dayside and nightside spectra.
We also perform a simple check on the spectral difference between evening and morning spectra.
We find that the maximum absolute spectral difference between the two spectra is about 2-$\sigma$ and the median value of the absolute spectral difference across different wavelengths is about 0.5$\sigma$, where $\sigma$ is the propagated observational uncertainties.
In conclusion, we do not find significant differences between the evening and morning spectra.
In our following analysis, we focus on analyzing the day(noon)-night(midnight) spectral variations.

\begin{figure}[!tbp]
  \centering
    \includegraphics[width=.5\textwidth]{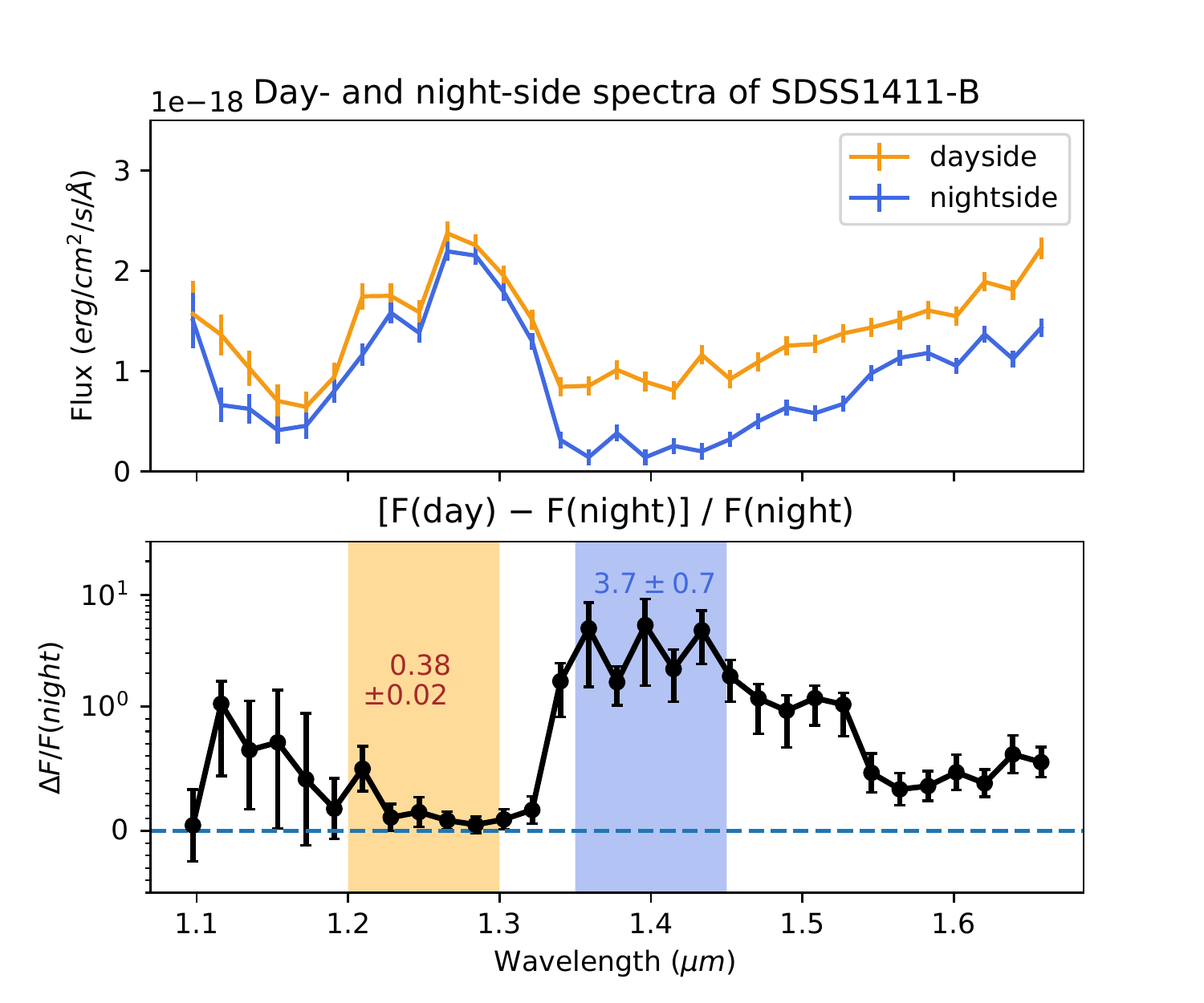}
    \caption{\textit{Top panel}: the binned dayside and nightside spectra of SDSS1411-B. The water-band flux of the dayside spectrum is almost two times higher than that of nightside spectrum.  \textit{Bottom panel}: the relative spectral variation between the dayside and nightside spectra. The spectral variation is almost ten times larger in the water band than in the $J'$ band. To show order-of-magnitude variations with uncertainty ranges that include negative values, we use the `symlog' y-axis scale, which is in linear scale below unity and in log scale above unity.} \label{fig:maxmin}
\end{figure}

 In the top panel of Figure~\ref{fig:maxmin}, we plot the extracted brown dwarf dayside and nightside spectrum. We find that the water-band flux in the dayside spectrum is almost twice as high as in the nightside spectrum.
In the bottom panel of Figure~\ref{fig:maxmin}, we plot the day-night spectral variations relative to the nightside spectrum.
We find that one of the key features in the relative spectral variation is the relatively low flux difference in the $J'$ band (1.2 -- 1.3\um). 
The binned $J'$-band flux difference is $38\pm2\%$, while the binned water-band flux difference is $370 \pm 70\%$.

\subsection{Brightness temperature variation}\label{sec:bt}
Brightness temperature is the temperature at which a blackbody emits the same amount of specific intensity as the observed value.
In an atmosphere where the temperature monotonically decreases with lower pressure, a hotter brightness temperature indicates that the flux emits from a higher pressure.
Based on the assumption of the monotonic relationship between temperature and pressure, brightness temperature is then a useful proxy to compare the pressure regions that different wavelengths probe.

\begin{figure}[!tbp]
    \includegraphics[width=.5\textwidth]{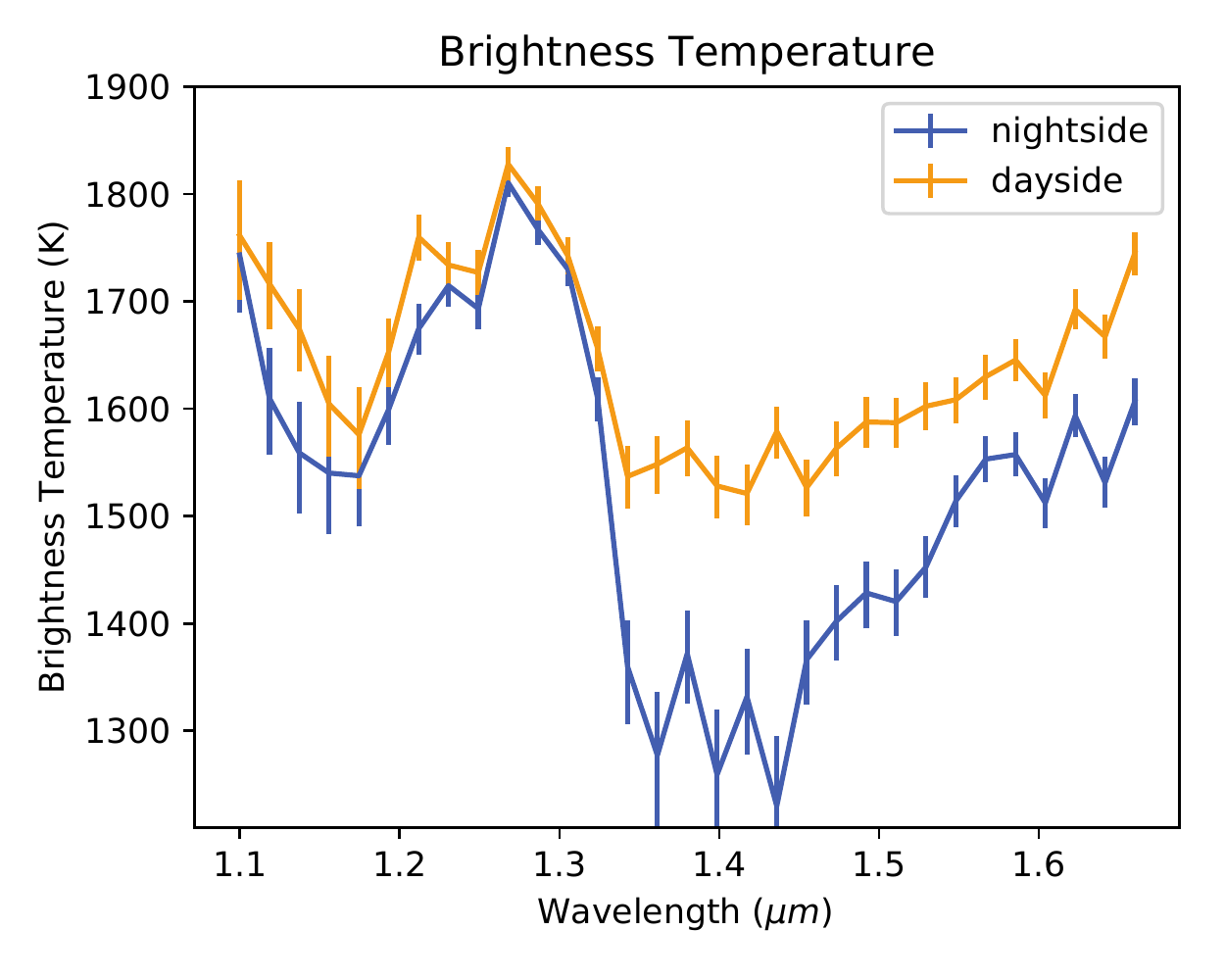}
    \caption{The brightness temperatures of the dayside and nightside spectra. The brightness temperatures are binned with a bin size of 0.014\um. The  brightness temperatures are higher in the $J'$-band than in the water bands.}
    \label{fig:bt}
\end{figure}

After converting flux densities to brightness temperatures with Planck equation and the derived brown-dwarf radius in Section \ref{sec:bdradius}, in Figure~\ref{fig:bt} we plot the brightness temperatures as a function of wavelengths for the dayside and nightside spectra.
 Based on Figure~\ref{fig:bt}, we conclude that the high-altitude atmosphere -- that emits the bulk of the water-band flux -- experiences a larger change in temperature than the low-altitude atmosphere that the J-band flux probes. 
 The conclusion that lower-altitude atmosphere has lower change in temperature variation is also consistent with the reported broadband photometric phase curve observation by \citet{casewell2018}.
  \citet{casewell2018} reported that the brightness temperature variations (360$\pm80$K) in the $K_{\rm s}$ band is higher than that (93$\pm$12K) in the $H$ bands.

\section{Modeling day-night spectral variations}\label{sec:models}

\subsection{Description of models}
To study the difference in atmospheric structure between the dayside and nightside atmosphere, we model the irradiated SDSS1411-B atmosphere with one-dimensional cloudless atmospheric models.
The models are based on \citet{mckay1989, marley1999, mayorga2019} with a wavelength grid ranging from 0.3 to 200\um.
The models assume a surface gravity of log(g) = 4.4, solar metallicity, and chemical-equilibrium abundances. 
The models properly account for the scattering and deposition of incident flux such that the irradiated atmospheres have non-zero albedo values.
The modeled dayside atmosphere is irradiated by a 13,000\,K white-dwarf at 0.003 au  \citep{littlefair2014}.
The shortest wavelength bin of our models is around 0.3~\um.
To include the irradiation flux at wavelengths shorter than 0.3\um, we increase the irradiation flux at the 0.3\um bin such that it is equivalent to the sum of the white dwarf irradiation flux at $\lambda < 0.3$\um, which is about 60\% of bolometric luminosity. 
This approximate treatment is necessitated by a lack of appropriate UV opacities required for the models.

\begin{figure}[!h]
    \centering
    \includegraphics[width=.5\textwidth]{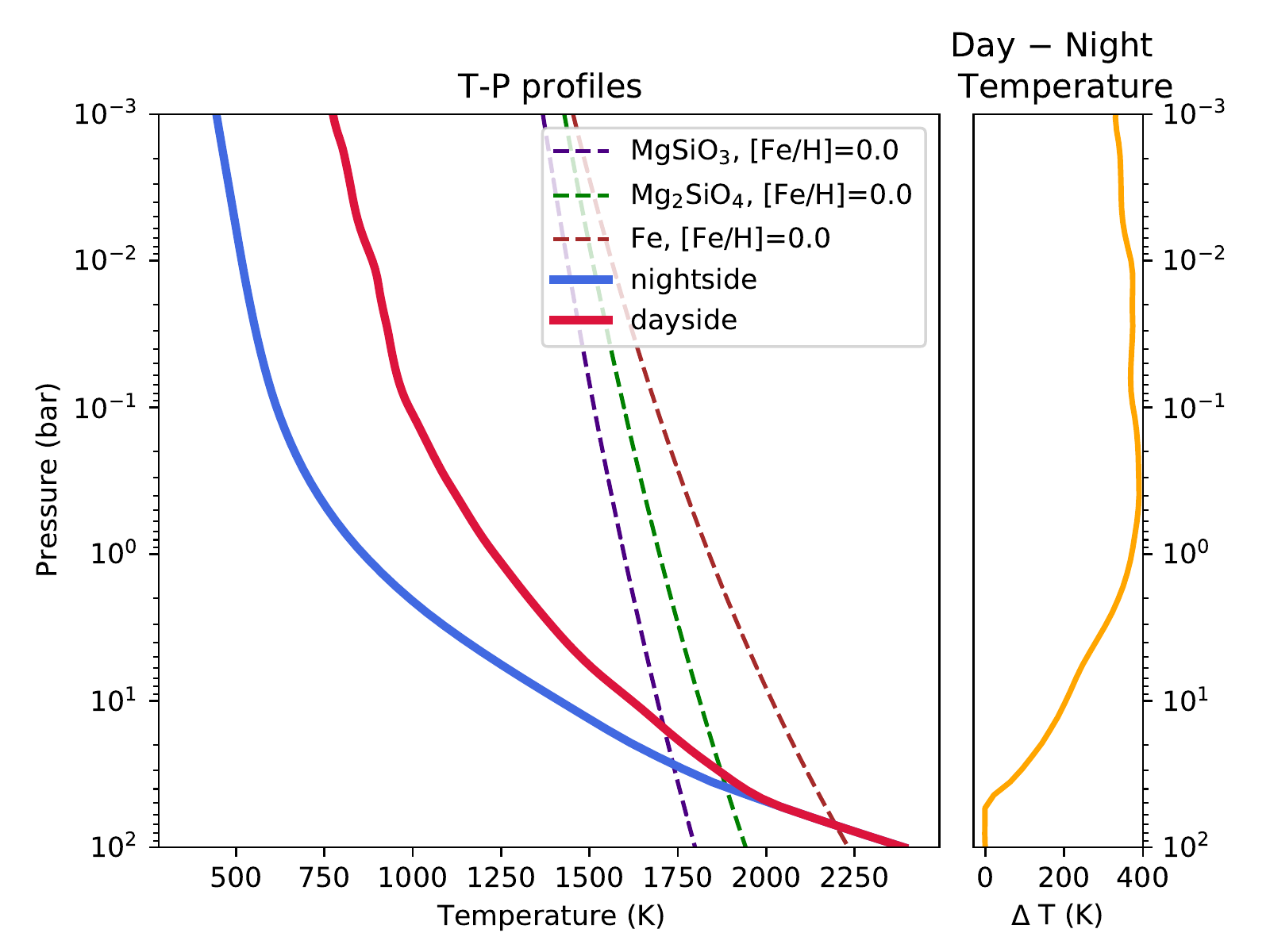}
    \caption{Left panel: the T-P profiles of nightside (solid blue line) and dayside (solid red line) models. The purple, green, and red dashed lines show the condensation curves of $\mathrm{MgSiO_3}$, $\mathrm{Mg_2 SiO_4}$, and Fe respectively from \citet{visscher2010}. The radiative-convective boundaries of nightside and dayside models are at around 30--60\, bars. Right panel: the day-night temperature difference.}
    \label{fig:pt}
\end{figure}

Further, we assume that the nightside T-P profile has an identical interior temperature (T(P=100\,bar)) to that of the dayside.
The modeled brown dwarf atmospheres are therefore have non-zero internal energy.
Because our non-irradiated model is designed to converge to a specified effective temperature, we construct two models whose effective temperatures are slightly lower and higher than that of the dayside model. 
We then linearly interpolate the two T-P profiles ($T_{\rm night1}(p)$ and $T_{\rm night2}(p)$) so that the interpolated T-P profile shares the same temperature with the dayside model at the pressure of 100 bars.
In the linear interpolation, we calculate the averaged temperature difference at P $>$90 bars between the nightside T-P profiles and the dayside T-P profile:
\begin{align}
\shortintertext{at p$>$90\, bars,}
\Delta T_1 &= \mid{\rm mean}{(T_{\rm day}(p) - T_{\rm night1}(p))}\mid\\
\Delta T_2 &= \mid{\rm mean}(T_{day}(p) - T_{\rm night2}(p)) \mid
\end{align}
We then assign a relative weighting for the two nightside T-P profiles based on the temperature differences at the high-pressure region for the linear temperature interpolation over isobars and obtain the interpolated nightside T-P profile ($T_{\rm newnight}(p)$):
\begin{align}
W &= \Delta T_1 + \Delta T_2, \\
{T_{\rm newnight}(p)} &= T_{\rm night1}(p) \times \frac{\Delta T_2}{W} + 
T_{\rm night2}(p) \times \frac{\Delta T_1}{W}
\end{align}
We use the PICASO model \citep{batalha2019} to calculate the outgoing emission based on the interpolated nightside T-P profile.
We plot the modeled dayside and nightside temperature-pressure profiles in Figure \ref{fig:pt}.

\subsection{Spectral modeling results}

\begin{figure}[!h]
    \centering
    \hspace{-0.75cm}
    \includegraphics[width=0.5\textwidth]{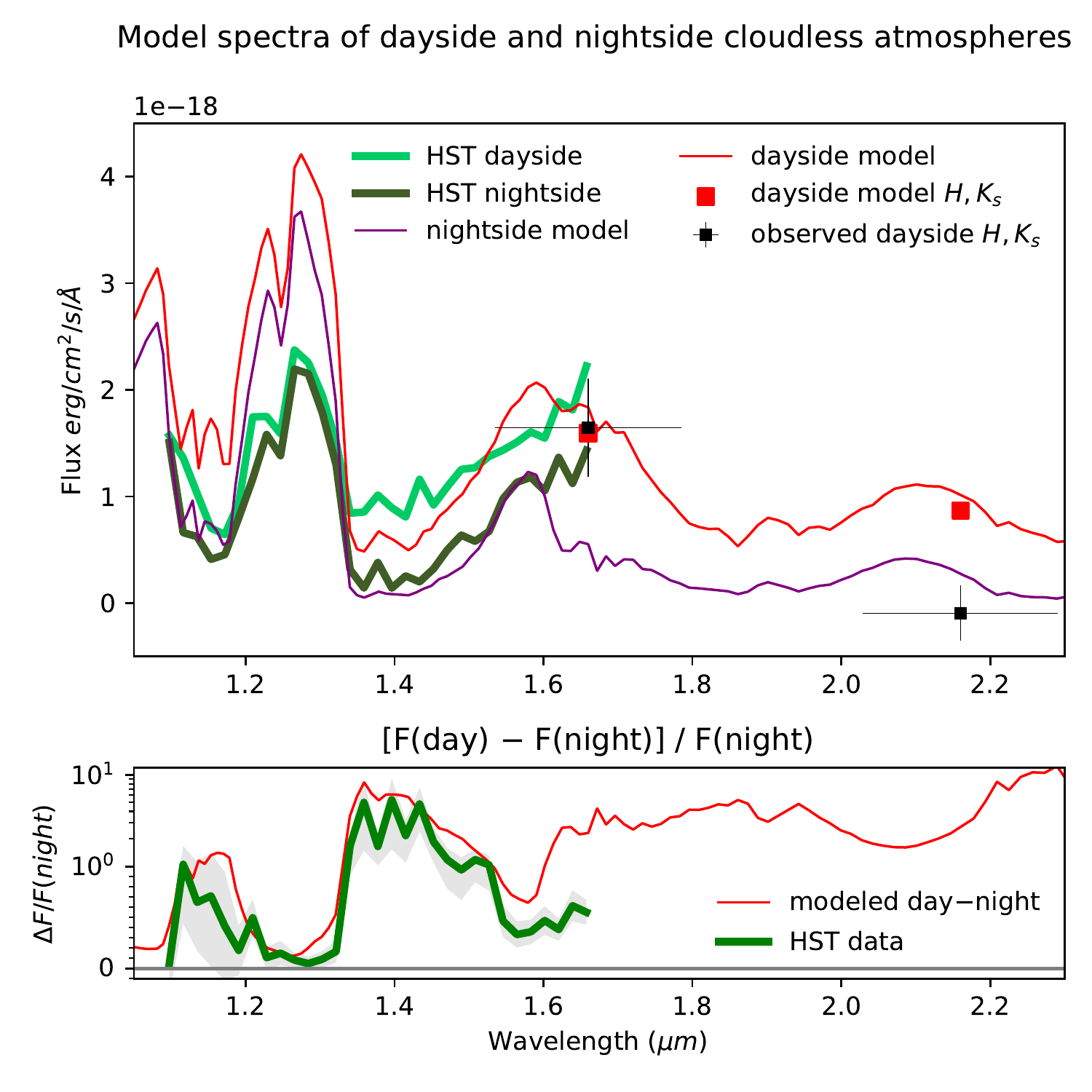}
    \caption{\textit{Top panel:} The nightside (purple line) and dayside (red line) model spectra  qualitatively reproduce the spectral shape of the observed nightside (dark-green line)  and dayside spectra (light-green line). We plot the ground-based $H$ and $K_s$-band photometry from \citet{casewell2018} as black squares at 1.66 and 2.16\um respectively. The modeled dayside $H$-band flux (red square) are consistent with observed values while the modeled $K_s$-band flux (red square) is higher than the observed value. See text for the discussion of the difference between modeled and observed spectra.
    \textit{Bottom panel:} the modeled day-night  spectral variation shows that the flux variation is about order-of-magnitude higher in the water-band than that in the J-band region, matching the observed wavelength-dependence in day-night spectral variation.
    }
    \label{fig:modelspec} 
\end{figure}

In Figure \ref{fig:modelspec}, we plot the modeled dayside and nightside spectra and the day-night spectral variation.
In addition to the HST WFC3/G141 spectra, we also compare our models with the ground-based $H$- and $K_s$-band photometry (black squares) from \citet{casewell2018}.
Based on the reported SDSS1411 WD+BD system magnitudes in \citet{casewell2018}, in Section \ref{sec:night} we subtract the best-fit white-dwarf model flux contribution  to derive the brown-dwarf dayside $H$- and $K_s$-band flux.
We find that the dayside and nightside atmospheric model spectra qualitatively reproduce the main spectral features, including the water band absorption at 1.15 and 1.45\um and the Potassium line (K I) absorption at around 1.27\um. 
Both the dayside and nightside models over-estimate the J-band flux and under-estimate the 1.4--1.67\um flux.
The ground-based $H$-band flux is consistent with the dayside-model flux within $1 \sigma$; the dayside-model $K_s$-band flux is about 4-$\sigma$ higher than the observed value, even though the discrepancy is sensitive to the derived white-dwarf  flux contribution (see Section \ref{sec:night}) and the  $K_s$-band absolute flux calibration.
The deviations between the data and our forward-model-based spectral fits suggest that our understanding of the atmospheric processes and line lists are incomplete for irradiated atmospheres. 
For example, absorption by strong UV transitions could heat up the pressure region probed by water and $H$-band flux and, therefore, increase the emission at these wavelengths. 
Furthermore, including cloud opacity could lower the modeled $J$-band flux and provide a better fit to the observed value. We discuss further the impact of potential clouds in Section \ref{sec:clouds}.

\subsection{Pressure-dependent temperature contrast}\label{sec:contrast}

 Constraining the pressures from which the flux at different wavelengths are emitted  is essential for mapping the observed spectral variation to the day-night temperature contrast as a function of pressure.
We calculate the contribution function of the nightside model to measure the relative flux contribution  to the top-of-atmosphere emission at each wavelength bin across different pressures.
To calculate the contribution function, we perturb the nightside T-P profile by increasing the temperature in each pressure layer by 100\,K and calculate the change in the top-of-atmosphere emission.
At each wavelength bin, we divide the emission change due to the temperature perturbation of each layer by the total emission change to obtain the relative flux contribution per layer.
 In Figure~\ref{fig:cf} we plot the contribution function for the nightside model spectrum.
The contribution function suggests that $J'$-band and water-band emission originates from around 20-80 bars and 2-20 bars region respectively.

With the calculated contribution function and the measured day-night brightness temperature contrast in Section \ref{sec:bt}, we illustrate the pressure-dependent day-night temperature contrast in Figure \ref{fig:pressdtemp}.
As shown in the right panel of Figure \ref{fig:pressdtemp}, the day-night temperature contrast increases with lower pressure.
We can qualitatively explain the trend in the pressure-dependent temperature contrast with cooling and advection timescales.
The radiative cooling timescale decreases with higher pressure \citep[e.g.,][]{showman2002}.
Assuming the advection timescale of atmospheric jets is roughly the same in the 1-50 bars range (see Figure 5 in \citealt{tan2020} and Figure 4 in \citealt{lee2020} for example), the ratio of radiative cooling over advection timescale decreases with lower pressure.
Therefore, the dayside atmosphere at a lower pressure cools faster, so the day-night temperature contrast is higher at lower pressure.
Besides the advection and cooling processes, convection could also play an important role in explaining the similar dayside and nightside temperatures, or low day-night temperature contrasts, at high pressure region.
Our atmospheric models indicate that the radiative-convective boundary (RCB) is at around 30--60 bars.
The interiors of brown dwarfs should be fully convective and the interior specific entropy is expected to be homogenized. 
At atmospheric layers deeper than the RCB, convection is expected to drive the profiles towards having the same specific entropy as the interior. 
Therefore, the dayside and nightside flux that originates below the RCB could be similar to each other because of convection.
Our results demonstrate the observational constraints on the pressure-dependent day-night temperature differences that extends to the high pressure region (P$>$20bars) of irradiated atmosphere.

\subsection{Temperature gradient}
\begin{figure*}
    \hspace{-1cm}
    \includegraphics[width=1.1\textwidth]{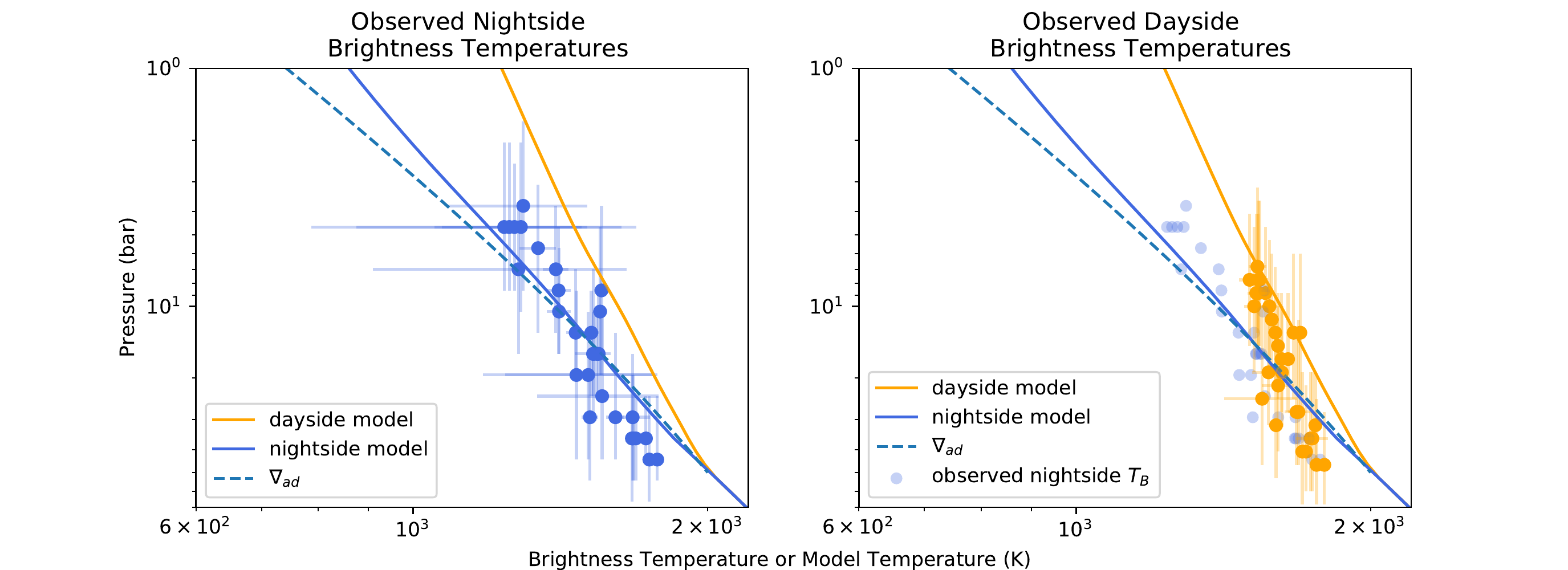}
    \caption{Left panel: The observed nightside brightness temperatures at different inferred pressures, which are derived from the contribution function of the nightside model, are plotted as blue points. The solid blue and orange lines show the modeled dayside and nightside temperature-pressure profile. The blue dashed line shows an adiabat profile that share the same interior temperature at 30\,bars with model. Right panel: Observed dayside brightness temperatures over different inferred pressures, which are derived from the contribution function of the dayside model, are plotted in orange points. The solid and dashed lines are the same as those in the left panel.}
    \label{fig:tempgrad}
\end{figure*}

In Figure \ref{fig:tempgrad}, we plot the observed brightness temperature over the inferred pressure based on the calculated contribution functions in Section \ref{sec:contrast}.
We emphasize that the inferred pressures are model-dependent while the brightness temperatures are derived from the observed flux densities.
We also plot the adiabat profile assuming the temperature gradient follows the adiabatic gradient derived by \citet{parmentier2015} based on the fit to the equation of state of substellar objects at high-pressure region  \citep{saumon1995}:
\begin{equation}
    \nabla_{ad} \approx 0.32 - 0.1 (\frac{T}{3000 K})
\end{equation}
We integrate the temperature gradient with a boundary condition such that the adiabatic profile shares the same temperature at high pressures ($\sim$ 50\,bars) with that of both the dayside and nightside models.
Both dayside and nightside brightness temperatures show a shallower temperature gradient than that of the adiabat (i.e., temperature decreases less with lower pressure than that of an adiabatic profile or $\mid$d$\log$T/d$\log$P$\mid < \mid\nabla_{\rm ad}\mid$), although the uncertainties are large.
In the right panel of Figure \ref{fig:tempgrad}, the dayside atmosphere has a shallower temperature gradient than that of the nightside atmosphere.

The thermal profile in irradiated objects is determined by the competition between the bottom heating via the internal flux and the top heating from the stellar light. When the received stellar irradiation is larger than the internal flux, it stabilizes the atmosphere against convection and reduces the thermal gradient. The effect happens mainly on the dayside, where the stellar heat is deposited, but it also happens on the nightside because of the advection of heat from day to nightside. Because this heat advection is not perfectly efficient, the nightside profile is closer to the adiabatic gradient than the dayside atmosphere.

\begin{figure}[!h]
    \centering
    \includegraphics[width=0.5\textwidth]{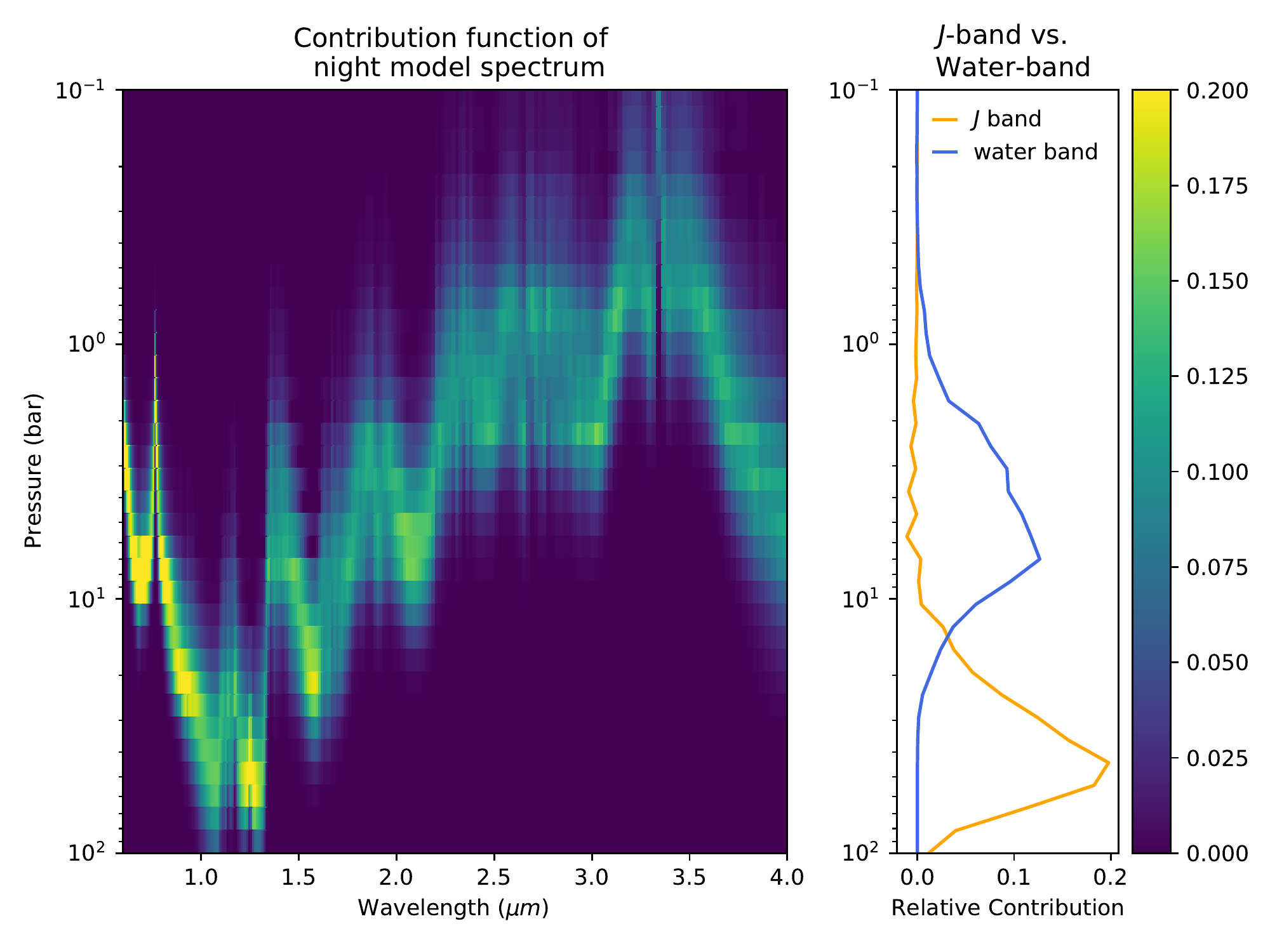}
    \caption{Left panel: the contribution function of the nightside model. The colors represent the relative contribution per pressure layer to the emission. Right panel: The peak of J-band and water-band contribution function of the nightside spectrum is at $\sim$20--80 bars and 2--20 bars respectively.}
    \label{fig:cf}
\end{figure}

\section{Discussion}\label{sec:discussion}

\begin{figure*}[!hbtp]
    {\hspace{-1cm}\includegraphics[width = 1.15 \textwidth]{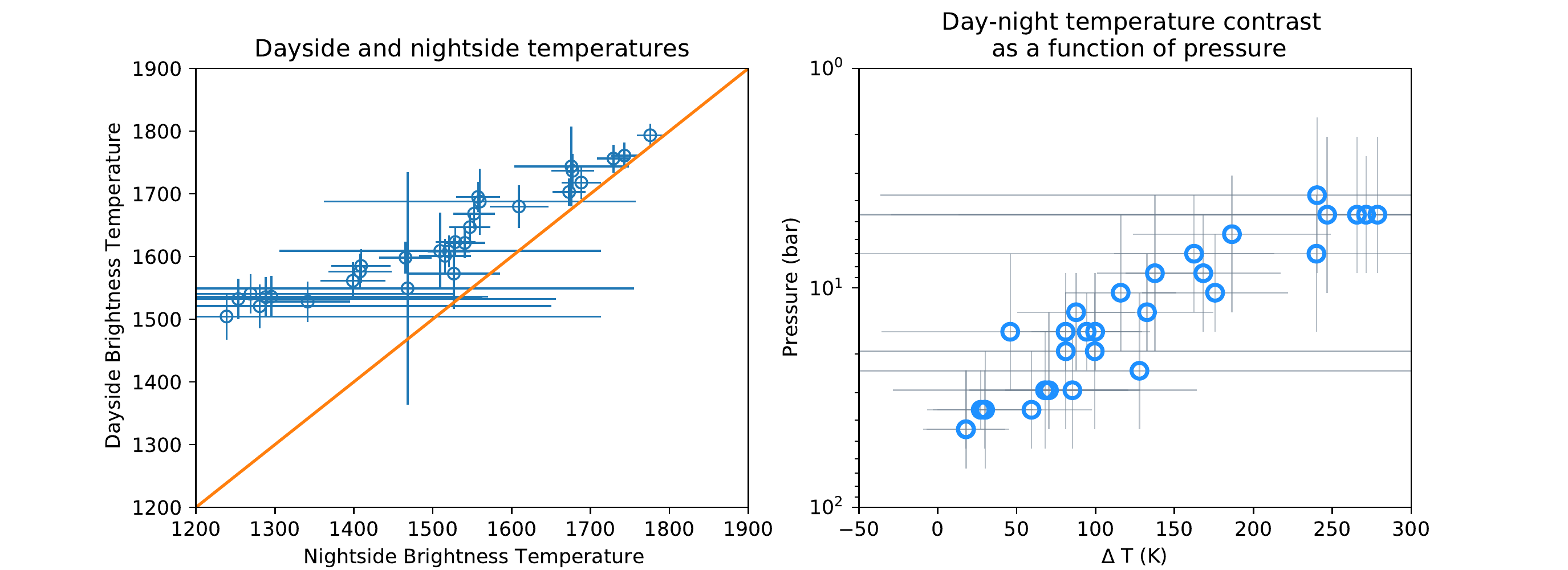}}
    \begin{minipage}{1.0\textwidth}
 \caption{Left panel: The nightside and dayside brightness temperatures are less different at higher nightside brightness temperatures. The orange y=x line portrays equal dayside and nightside temperatures. Right panel: The measured day-night brightness temperature contrast as a function of nightside pressures suggests that day-night temperature contrast increases with lower pressure.}    \label{fig:pressdtemp}
 \end{minipage}
\end{figure*}

  \subsection{Comparison of Spectra and Color Variations to those in isolated brown dwarfs}
\begin{figure}[!h]
    \hspace{-0.5cm}
    \includegraphics[width=0.54\textwidth]{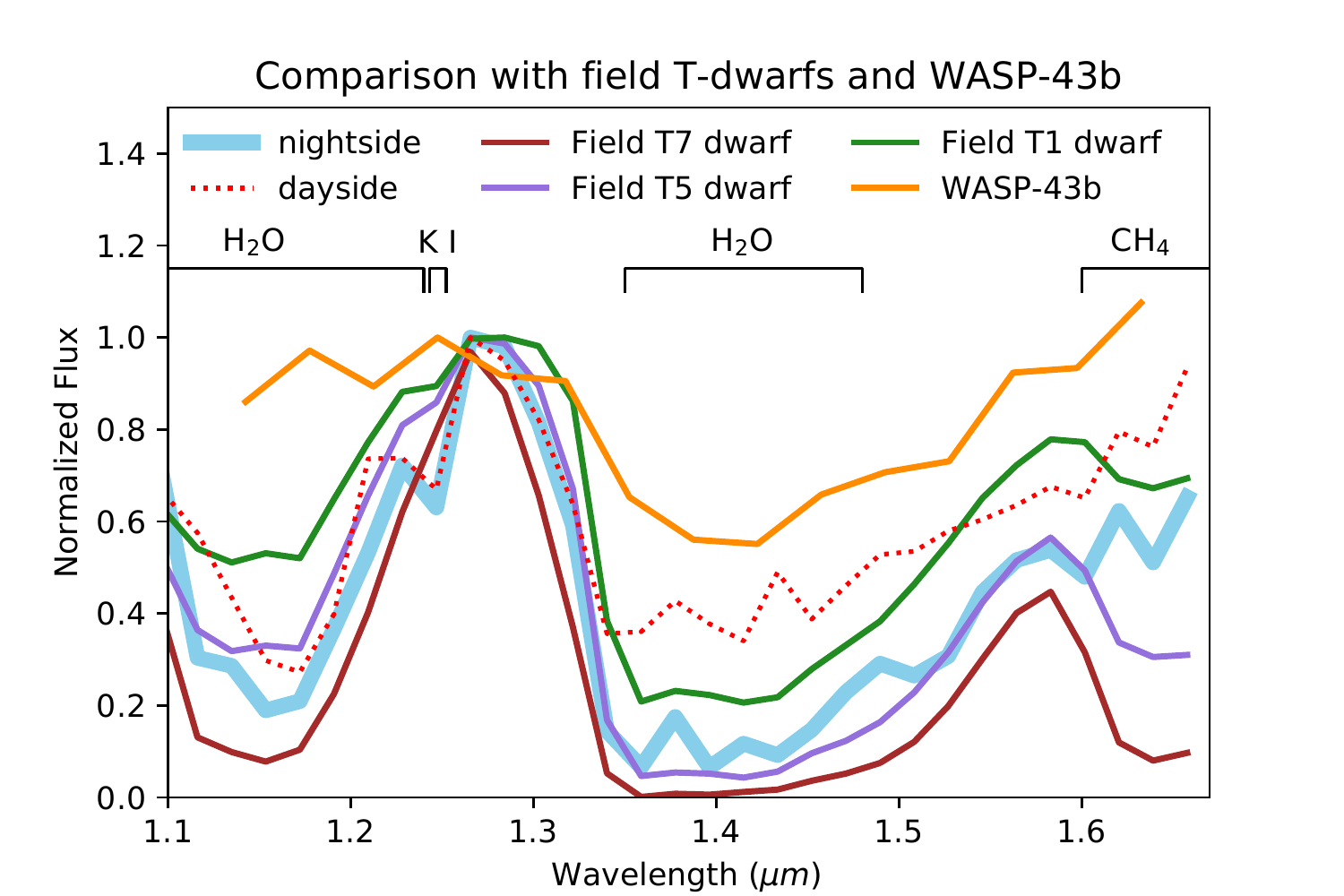}
    \caption{Comparison of the binned dayside (dotted red line) and nightside (solid blue line) SDSS1411-B spectra with field T-dwarfs and WASP-43\,b spectra. The SDSS1411-B nightside spectrum (thick light-blue line) shows a similar water-band absorption depth with the field T5 dwarf (solid purple line) and lack of strong methane absorption feature at 1.65\um which is prevalent among T dwarf spectra. The higher gravity and the steeper temperature gradient of SDSS1411-B causes the deeper water-band absorption compared to that of the nightside spectrum of hot Jupiter WASP-43\,b (solid orange line). We also plot the spectra of field T1 and T7 dwarfs in solid green and maroon lines respectively for reference.}
    \label{fig:fieldspec}
\end{figure}
How does the redistribution of irradiation energy from the dayside to the nightside affect the nightside emission spectra of the irradiated brown dwarf SDSS1411-B?
In Figure~\ref{fig:fieldspec}, we compare the spectra of a field T1 dwarf (2MASS  J10393137+3256263, \citet{buenzli2014,manjavacas2019a}), field T5 dwarf (2MASS  J00001354+2554180, \citealt{buenzli2014,manjavacas2019a}), and field T7 dwarf (2MASSI J1553022+153236, \citet{burgasser2010})  with that of SDSS1411-B, which are normalized by the flux at around 1.27\um.
SDSS1411-B nightside emission spectra show a similar water-band absorption depth to that of the field T5 dwarf, which has an effective temperature of around 1000\,K \citep{filippazzo2015}.
However, we notice the lack of methane absorption feature in the nightside emission spectra at around 1.6--1.67\um, which is prominent in the spectra of T dwarfs.  
We speculate that the shallower methane absorption feature in the nightside emission spectra is likely because SDSS1411-B is hotter than field T dwarfs in the low pressure region probed by H-band flux, which is approximately about 10 bars as indicated by our contribution function in Figure \ref{fig:cf}.
It is also possible that the global atmospheric circulation drives the methane abundance out of chemical equilibrium, a process that has been studied for hot Jupiter atmospheres \citep[e.g.,][]{cooper2006,agundez2014b,steinrueck2019a,drummond2020}.
Finally photochemical process could destroy $\rm{CH_4}$ in the dayside atmosphere \citep{zahnle2014}.
Further studies into how the atmospheric circulation in such rapidly rotation irradiated atmospheres affects horizontal and vertical quenching are needed to fully understand the impact of atmospheric dynamics on molecular abundances.

\begin{figure}
    \centering
    \includegraphics[width=0.43\textwidth]{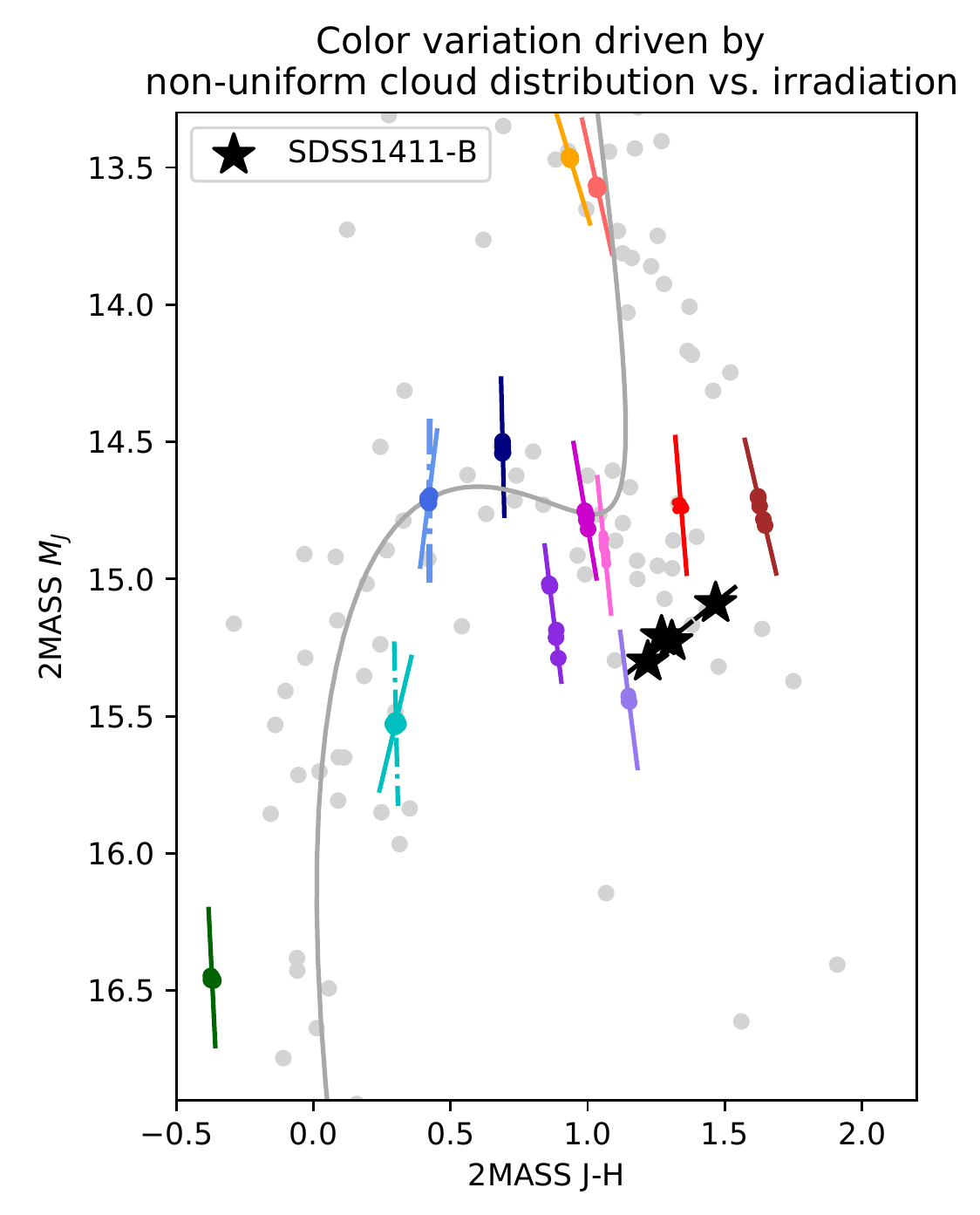}
    \caption{The isolated brown dwarfs, plotted as colored round dots, manifest weak color variations in rotational modulations \citep{lew2020a}; SDSS1411-B, plotted as black stars, show significant change in $J-H$ colors as the irradiation varies across the night (faintest $M_J$), morning, evening, and noon (brightest $M_J$) orbital phases. The colored solid lines are the fitted straight lines to the color-magnitude variations of each objects. The dash-dotted lines show the semi-major axis of the fitted ellipses to the color-magnitude variations of two T dwarfs, GU Piscium b and 2MASS J22282889-431026. The grey colored dots show the $J-H$ colors and $J$-band magnitudes of field L and T dwarfs based on the catalog managed by \citet{dupuy2012}. The grey curve shows the evolution from L (upper right) to T spectral type (bottom left).}
    \label{fig:cmd}
\end{figure}

The observed color variations of SDSS1411-B are mainly driven by the orbital phase-dependent irradiation and atmospheric circulation.
In contrast, the observed color variations of isolated brown dwarfs are likely driven by the heterogeneous clouds in their atmosphere.
In Figure \ref{fig:cmd}, we compare the 2MASS $J-H$ color variations of  SDSS1411-B with that of twelve isolated L-to-T dwarfs, as reported in \citet{lew2020a}.
The plot is the same as Figure 6 in \citet{lew2020a} with the addition of SDSS1411-B colors at four orbital phases (i.e., the four spectra in Figure \ref{fig:phasespec}). 
All plotted objects are observed under the HST/WFC3/G141 mode.
Each colored datapoint represents the approximated $J-H$ colors and absolute magnitudes $M_J$ averaged over one HST orbit
of each isolated brown dwarf.
The solid lines indicate, for each object, the linear extrapolation for its $\Delta J~ vs. \Delta(J-H)$ trend.
Because the WFC3/G141 spectra do not fully cover the 2MASS $H$-band, we derive the 2MASS $H$ magnitude of SDSS1411-B with the white-dwarf brown-dwarf $H$-band magnitude (17.80$\pm 0.04$) from \citet{casewell2018} and the best-fit white-dwarf model in Section \ref{sec:night}.
We subtract the $H$-band magnitude of 17.99 of the best-fit white-dwarf model from 17.80 to obtain the brown dwarf $H$-band magnitude of 19.6$\pm 0.4$.
Finally, we also assume the $H$-band magnitude variations are the same as that in the partial $H$-band coverage of the WFC3/G141 spectra.

Based on Figure \ref{fig:cmd}, it is clear that no strong color variations are found among the L and T dwarf atmospheres, as concluded in \citet{lew2020a}.
In contrast, we observe strong $J-H$ color changes (compared to its J-band brightness variation) for SDSS1411-B.
The comparison of color-magnitude variations between SDSS1411-B and isolated brown dwarfs demonstrate the impact of different atmospheric processes on the near-IR colors in brown dwarf atmospheres.

\subsection{Comparison to hot Jupiter WASP-43 b}
SDSS1411-B shares a similar irradiation temperature ($T_{\rm irr}$ $\sim$ 1300K with zero albedo) with many hot Jupiters but has a higher mass and internal energy budget.
In Figure \ref{fig:fieldspec}, we compare the spectral features of SDSS1411-B to that of hot Jupiter WASP-43\,b \citep{stevenson2014}, which has an equilibrium temperature of around 1440 K \citep{blecic2014} and is one of the few hot Jupiters with a measured spectroscopic phase curve.
A key difference in the nightside emission spectra between the two strongly irradiated atmospheres is the water-band feature -- SDSS1411-B has a much deeper water-band absorption than that of WASP-43\,b.

We attribute the difference in the water-band flux between SDSS1411-B and WASP-43\,b to the difference in their interior energy and gravity.
Brown dwarf evolutionary model Sonora \citep{marley2018} estimates an effective temperature of around 900\, K for an isolated brown dwarf that shares the same mass (50 $M_{\rm Jup}$ and age ($>$3 Gyr) with SDSS1411-B.
 In contrast, typical gas giant planets with an age of a few Gyrs old like Jupiter are often assumed to have an intrinsic temperature, which is the effective temperature in the absence of irradiation, of around 100--250\,K \citep{fortney2008}.
 
 The interior energy is important for the emission spectral feature because it affects the RCB and the corresponding temperature gradient.
 At the same gravity, an atmosphere with a higher interior energy has a RCB at a lower pressure.
With a lower RCB pressure, the adiabatic temperature gradient in the convection zone, which is steeper than the temperature gradient in the radiative zone, extends to a lower pressure.
Flux originated from the part of atmosphere with the steeper temperature gradient manifests a stronger spectral feature such as water-band absorption.
 Therefore, the higher internal energy of SDSS1411-B than WASP-43b leads a lower RCB pressure and thus shows a stronger emission spectral features including the water-band absorption.
 
In addition to the interior energy, the gravity is important to understand the spectral feature difference because gravity affects the photospheric pressure.
 SDSS1411-B (M$\sim 50 M_{\mathrm{Jup}}$) is about 25 times more massive than that of WASP-43\,b ($\sim 2M_{\text{Jup}}$) with a similar radius.
The photospheric pressure $P$ at $\tau =1$ is proportional to the gravity $g$, i.e., $P(\tau=1) \propto g$. 
With the higher gravity, the near-infrared emission probes a higher pressure which is closer to or deeper into the convection zone.
As mentioned earlier, the temperature gradient is steeper in the convection zone than in the radiative zone.
The likely steeper temperature gradient combined with the higher photospheric pressure results in a deeper water-band absorption in the atmosphere of SDSS1411-B than in that of WASP-43\,b.
We note that the effects of pressure broadening should also increase with a higher photospheric pressure. 
Pressure broadening typically weakens spectral features, and thus partially cancels out the impact of a high internal energy flux and a high photospheric pressure, which would both work to strengthen the spectral features, such as the water-band absorption feature we discuss here.

Besides the difference in interior energy and gravity, the irradiation luminosity of WASP-43\,b is about 1.5 times higher than that of SDSS1411-B. 
The striking difference in the host star effective temperature (4,400\,K vs 13,000\,K) means that the majority of stellar flux received by WASP-43\,b is in optical wavelength range, and thus the subsequent heating of atmosphere depends on the optical opacities;
In contrast,  SDSS1411-B mainly receives irradiation flux at UV wavelengths, so UV opacity becomes critical for atmospheric heating.
 Spectral or photometric observations of SDSS1411-B at longer wavelength that probe low pressure is vital for studying the T-P profile and atmospheric composition under strong UV irradiation.

\subsection{Day-night temperature contrast, the nearly zero phase offset, and implications for atmospheric dynamics} 

Atmospheric circulation models \citep[e.g., see][]{showman2020,lee2020,tan2020} for a rapidly rotating atmosphere under strong irradiation predict a large fractional day-night temperature contrast ratio, which is the ratio of the temperature contrast with irradiational heat redistribution over the radiative-equilibrium temperature contrast (see Figure 6 in \citealt{komacek2016} and Equation \ref{eq.eqdaynight}), and a small phase offset. Below we discuss the measured values of SDSS1411-B, and place them in the context of  close-in gas giants including hot Jupiters.

Assuming no heat redistribution, the dayside radiative equilibrium temperature ($ T_{\rm eff,day}$) is determined by both stellar irradiation and internal heat flux (which, in turn, is characterized by the intrinsic temperature $T_{\rm int}$, which is equal to the effective temperature of an isolated brown dwarf). In contrast, the nightside is purely set by the internal heat: 
\begin{equation}
\begin{split}
    & T_{\rm eff,day} = (T_{\rm int}^4 + T_{\rm irr}^4 )^{1/4},\\
    & T_{\rm eff,night} = T_{\rm int},\\
    & \Delta T_{\rm eff} = T_{\rm eff,day} - T_{\rm eff,night},\\
    \end{split}
\label{eq.eqdaynight}
\end{equation}
where $T_{\rm irr}$ is the dayside equilibrium temperature with zero albedo, global heat redistribution, and negligible interior energy budget and $\Delta T_{\rm eff}$ is the radiative-equilibrium temperature contrast.
We estimate the intrinsic temperature $T_{\rm int}$ of SDSS1411-B, to be around 900$\pm 200\,\rm K$ based on the Sonora brown dwarf evolutionary model, which does not include irradiation, for a $50 M_{\mathrm{Jup}}$ mass object with an age of 3-10 Gyr.
We note that the preceding common envelope evolutionary phase interaction between the brown dwarf and the white dwarf progenitor makes the deduction of the main sequence mass of the star, and hence its main sequence lifetime (and thus the evolutionary age of the brown dwarf) uncertain \citep[see, e.g.,][]{geier2011}. Any potential mass accretion or loss of the brown dwarf during the AGB/RGB phase of the star would add to this uncertainty \citep[see, e.g.,][]{lagos2021,villaver2007}.)

We adopt a surface gravity of $2500\,\rm ms^{-2}$ for the analytical model. 
We integrate the 1.1 -- 1.7\um spectra and calculate the WFC3 band-averaged day-night brightness-temperature difference of SDSS1411-B.

\begin{figure}
    \centering
    {\hspace{-0.5cm}
    \includegraphics[width=0.5\textwidth]{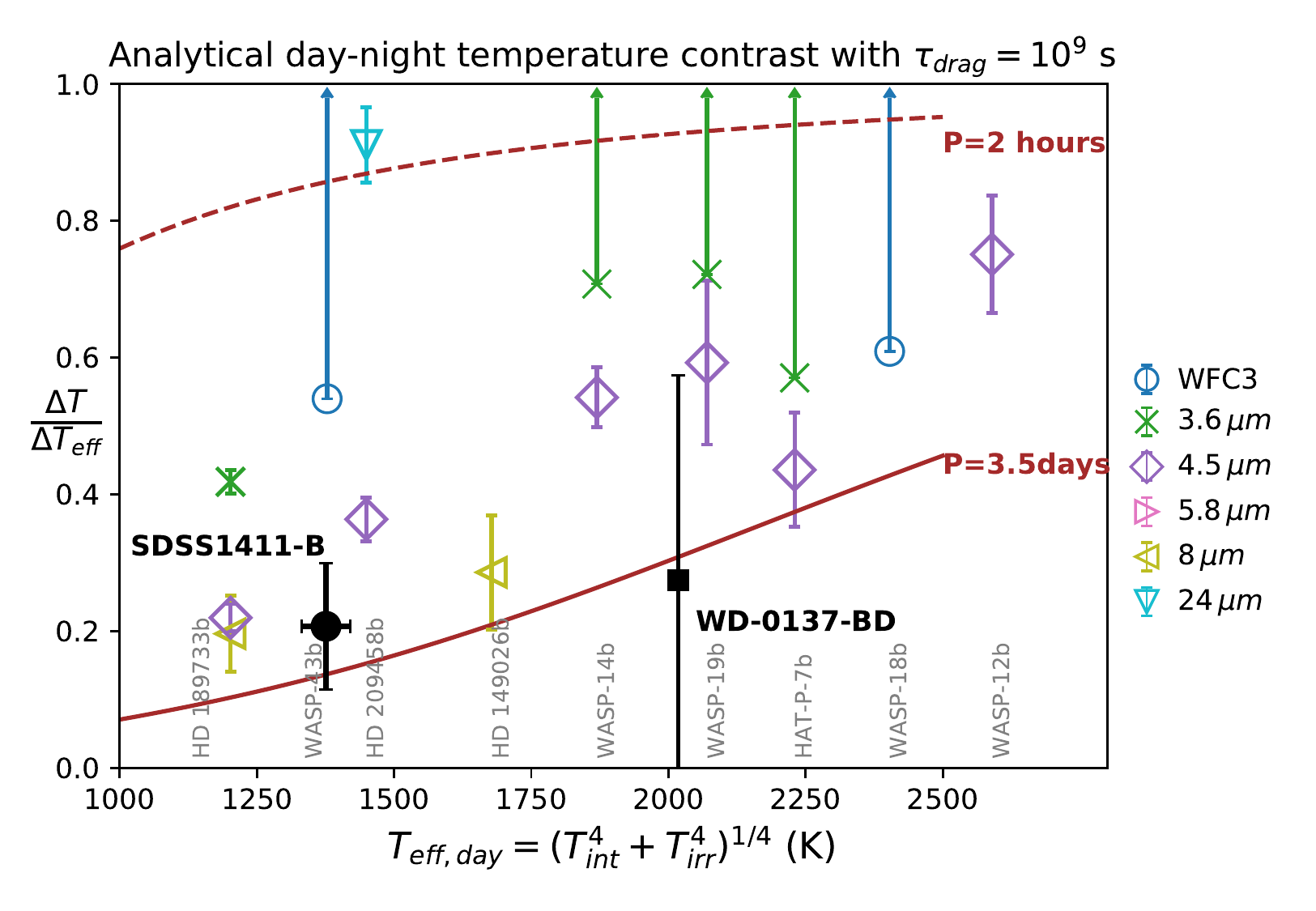}}
    \caption{The analytical estimate of day-night temperature contrast ratio with rotational periods of 2 hours  (dashed brown line) and 3.5 days (solid brown line) with a fixed drag timescale of $10^9$ seconds.
    The temperature contrast ratio of SDSS1411-B, shown in solid black circle, is lower than the analytical estimate with a period of two hours.
    The analytical temperature contrast ratio with the $10^9$ seconds drag timescale is a lower limit because the temperature contrast ratio increases with shorter drag timescale. 
    The colored points show the observed day-night temperature contrast ratio of hot Jupiters listed in Table 2 in \citet{komacek2016}, including HD 189733b \citep{knutson2009a,knutson2012}, WASP-43b \citep{stevenson2014}, HD 209458b \citep{crossfield2012,zellem2014}, HD 149026b \citep{knutson2009a}, WASP-14b \citep{wong2015}, WASP-19b \citep{wong2016}, HAT-P-7b \citep{wong2016}, WASP-18b \citep{arcangeli2019}, and WASP-12 b \citep{cowan2012}.
    The temperature contrast ratios of hot Jupiters observed at different wavelengths are plotted in different colors and markers.
    The observed temperature contrast ratio of WD-0137-BD \citep{lothringer2020}, which is another irradiated brown dwarf, is plotted as a black square.
    }
    \label{fig:komacek2016}
\end{figure}

The ratio between the measured day-night temperature difference and the estimated radiative equilibrium difference is shown in Figure \ref{fig:komacek2016} along with measurements of hot Jupiters as a function of equilibrium temperature. Theoretical predictions using the theory of \cite{komacek2016} with two planetary rotation periods (2 hours and 3.5 days)  and a negligible drag ($\tau_{\rm drag} = 10^9\, \rm s$)  are plotted as curves for comparisons. As shown in Figure \ref{fig:komacek2016}, the expected fractional temperature contrast ratio increases significantly with shorter rotational period (i.e, the dashed vs. solid brown lines) given the same equilibrium temperature. The measured value of SDSS1411-B is among those of hot Jupiters at nearby equilibrium temperatures which have a wide scatter. The measured temperature contrast ratio of SDSSJ1411-B is much lower than the analytical estimate at a rotation period of two hours.

Two factors may contribute to this discrepancy. First, the wavelength range in our study generally probes a deep atmosphere where the brightness temperature greatly exceeds $T_{\rm int}$. For example, the brightness temperature at the WFC3 bandpass ranges from 1000 K to over 1800 K with an average of $\sim1400$\,K based on the Sonora cloud-free models \citep{marley2018} with $T_{\rm int}=T_{\rm eff}=900$~K and g = 1780$\rm\,ms^{-2}$. This likely shifts the intrinsic temperature ``seen" by our measurements considerably higher than the assumed $T_{\rm int}$ based on the evolution models. With a higher ``$T_{\rm int}$" in Equations \ref{eq.eqdaynight} for our wavelength range, the inferred fractional day-night temperature contrast ratio can move closer to the theoretical expectation. Second, some processes may help to increase day-night heat transport and lower the expectation shown in Figure \ref{fig:komacek2016}. 
For example, if the latitudinal width of the standing waves and equatorial jet is larger than we expected, this could help to increase the day-to-night heat transport. One possible cause may be the increase of effective static stability of the atmosphere due to a strong dayside thermal inversion in the upper atmosphere (not probed by our HST observations, \citealt{lothringer2020}).

The lack of phase offset with respect to the expected secondary-eclipse phase (see Figure \ref{fig:centroid}) is consistent with the lack of a pronounced equatorial and/or zonal wind pattern in \citet{tan2020,lee2020}.
Further, the negligible phase offsets across multiple bands in Section \ref{sec:phaseoffset} hint that the horizontal thermal structure could be similar across a wide pressure range.

\subsection{Clouds} \label{sec:clouds}

Based on the T-P profiles and condensation curves of common cloud species plotted in Figure~\ref{fig:pt}, it is possible that clouds form at the near-infrared photosphere pressures of the dayside and nightside hemispheres.
While we defer the exploration of possible cloud structure to future studies, we herein provide a brief qualitative discussion of the potential impact of cloud formation in the irradiated brown dwarf atmosphere.

Given the same effective temperature, gravity, and irradiation, a cloudy atmosphere's model spectrum is fainter in the $J$-band and has weaker spectral features in comparison with a cloudless model spectrum.
With the addition of cloud opacity, the photospheric pressure, particularly that in the $J$-band in Section \ref{sec:contrast}, shifts to a lower pressure.
While the absolute photospheric pressure could shift in the presence of clouds, the relative pressure probed  at different wavelengths should remain roughly the same.
Therefore, we argue that the presented trend of higher temperature contrast with lower pressure shown in Figure \ref{fig:pressdtemp} remains valid in the cloudy atmosphere scenario. 
The estimated temperature gradient (Figure \ref{fig:tempgrad}) could be steeper in a cloudy atmosphere because cloud opacity limits the pressure range probed by the HST observation.

We can also qualitatively discuss the potential cloud coverage difference between the dayside and nightside atmospheres.
Cloud modeling studies for hot Jupiter atmospheres with equilibrium temperatures near 1400\,K \citep[see Figure 19 in][]{parmentier2021} show that the $J$-band brightness temperature difference between a fully cloudy nightside and cloudless dayside could be on the order of several hundred Kelvin.
We argue that the observed small ($\sim$ 10\%) day-night $J$-band brightness temperature variations (Figure \ref{fig:bt}), which is sensitive to cloud opacity, hints that it is unlikely that there is a drastic change in cloud coverage, such as a fully cloud-free to cloudy transition, between the dayside and nightside atmospheres.
Meanwhile, global cloudless atmospheric circulation studies \citep[e.g.,][]{tan2020,lee2020} of rapidly rotating irradiated atmospheres suggest a strong latidudinal-dependent temperature profile because of the fast rotation.
Further studies on the vertical mixing, latitudinal and longitudinal variations of T-P profiles, and the corresponding cloud structure are important to interpret the disk-integrated emission spectra of a cloudy atmosphere with rapid rotation.

\section{Conclusions}

We present the first time-resolved spectrophotometry of the white dwarf--brown dwarf close binary system SDSS J141126.20+200911.1, observed by the Hubble Space Telescope Wide Field Camera 3's G141 near-infrared grism. Our data baseline covers 8.666 hours or around 4.3 orbits of the system. 
Our high-precision data provides the first complete orbital-phase resolved spectroscopy of a highly irradiated brown dwarf and, as such, opens a new window on yet unexplored type of atmospheres. The key findings of our study are as follows:

\begin{enumerate}
    \item The broadband (1.15--1.67\um) white light curve, which includes flux from the white dwarf and the brown dwarf of SDSS1411 system, varies by around 2.8\% from trough to peak over an orbital period of 2.0287~hours.  
    The best-fit light curve model results show that the white dwarf broadband emission is about 12 times brighter than the time-averaged brown dwarf's emission.
    If only the brown dwarf's emission varies across orbital phases while the white-dwarf emission remains constant, the observed 2.8\% variability amplitude of the white dwarf-brown dwarf light curve implies that the brown dwarf broadband emission varies by around 38\% from the nightside to the dayside phase.
    \item A single sinusoidal fit to the light curve indicates that the phase offset, if any, is less than a 3-$\sigma$ upper limit of 11 degrees from the center of the secondary eclipse phase.
    \item We directly detect the nightside spectrum of SDSS1411-B during the eclipse of the white dwarf. We report a $J'$-band (1.2--1.3\um) detection at the 11$\sigma$ level and an $H'$-band (1.5--1.6\um) detection at the 7$\sigma$-level.
    \item We extract the brown-dwarf spectra at morning, noon, evening, and night phases. The flux variation between the dayside and nightside hemispheres of SDSS1411-B in the water band (1.35--1.45\um) is 370 $\pm 70$\%, which is about ten times higher than the 38$\pm 2\%$ variation observed in the $J'$ band. 

    \item Our atmospheric model spectrum reproduces the overall spectral feature of the observed nightside spectrum. Based on the nightside atmospheric model, the $J'$-band emission emerges from the 20--80 bar region. Based on the estimated $J'$-band photosphere pressures, we interpret that the day-night temperature contrast extends down to at least 20 bars and causes the observed $J'$-band flux variation.
        
    \item Based on the day-night brightness-temperature difference and nightside  atmosphere model, we calculate the pressure-dependent day-night temperature contrast. We show that the temperature contrast increases with lower pressure in the 2-80 bars pressure range that are probed by the HST 1.1-1.7\um  observation. The derived temperature gradient based on the brightness temperatures and inferred pressures suggest that the dayside atmosphere has a shallower temperature gradient (closer to an isothermal-like profile) than that of the nightside atmosphere.
    \item We show that the spectra of SDSS1411-B and of isolated T dwarfs manifest different spectral features. The color-magnitude variations of SDSS1411-B, which are driven by the varying irraditions, are also distinct from that of many isolated, rotating L and T dwarfs (driven by non-uniform cloud distributions). Our comparative study of spectra and color variations between irradiated brown dwarfs and isolated field dwarfs  illustrate the impact of irradiation on brown dwarf atmosphere. 
    
\item The comparison between SDSS1411-B and WASP-43\,b  nightside spectra demonstrates that a higher gravity and a higher internal heat flux lead to a stronger water-absorption feature. 

\item We find that the observed  day-night temperature contrast of SDSS1411-B is lower than the analytical estimate of \citet{komacek2016}. We interpret that this could be caused by the fact that 1.1.-1.67\um emission spectra probe atmospheric regions that are hotter than the $T_{\rm int}$ and/or a more effective day-night heat transport.
\end{enumerate}

Our observations successfully demonstrate that high-precision, orbital-phase resolved spectrophotometry of highly irradiated brown dwarfs is possible with the Hubble Space Telescope. Our analysis shows the potential for characterizing these atmospheres, the yet unexplored link between non-irradiated brown dwarfs and highly irradiated hot Jupiters. Analysis of our observations provides the first strong constraints on the details of the some constraints on atmospheric circulation and the pressure-temperature profiles of irradiated brown dwarfs.

\acknowledgments
We thank the anonymous referee for the careful review and constructive suggestions that significantly improve the paper.
B.W.P.L would like to acknowledge Dr. Adam Showman for his contribution to the observing proposal.
B.W.P.L thank Elsie Lee for the inspiring discussions during STScI Spring Symposium that helps improving the manuscript.All of the data presented in this paper were obtained from the Mikulski Archive for Space Telescopes (MAST) at the Space Telescope Science Institute. The specific observations analyzed can be accessed via \dataset[10.17909/t9-s253-4582]{https://doi.org/10.17909/t9-s253-4582}.
Support for Program number HST-GO-15947 and HST-AR-15060 was provided by NASA through a grant from the Space Telescope Science Institute, which is operated by the Association of Universities for Research in Astronomy, Incorporated, under NASA contract NAS5-26555.
This work is partly supported by the international Gemini Observatory, a program of NSF’s NOIRLab, which is managed by the Association of Universities for Research in Astronomy (AURA) under a cooperative agreement with the National Science Foundation, on behalf of the Gemini partnership of Argentina, Brazil, Canada, Chile, the Republic of Korea, and the United States of America.

%

\section*{Authors Contributions}
B.W.P.L reduced the data, performed the time-series and spectral analysis, and led the writing of the manuscript. 
D.A., M.M., S.X., and Y.Z. prepared the observing proposal.
S.C. provided the summary of the target properties. 
M.M and L.M. simulated the atmospheric model spectra.
X.T., V.P., L.M., and M.M aided in interpreting the atmospheric structure and circulation.
All authors contributed to the manuscript preparation.


\software{Astropy \citep{astropy:2013}, Numpy \citep{numpy}, Matplotlib \citep{matplotlib}, Scipy  \citep{scipy}, Batman \citep{kreidberg2015}}


\newpage
\appendix

\section{Validation of spectra extracted with the new 
pipeline}

\begin{figure}[!h]
    \centering
    \includegraphics[width=1.0\textwidth]{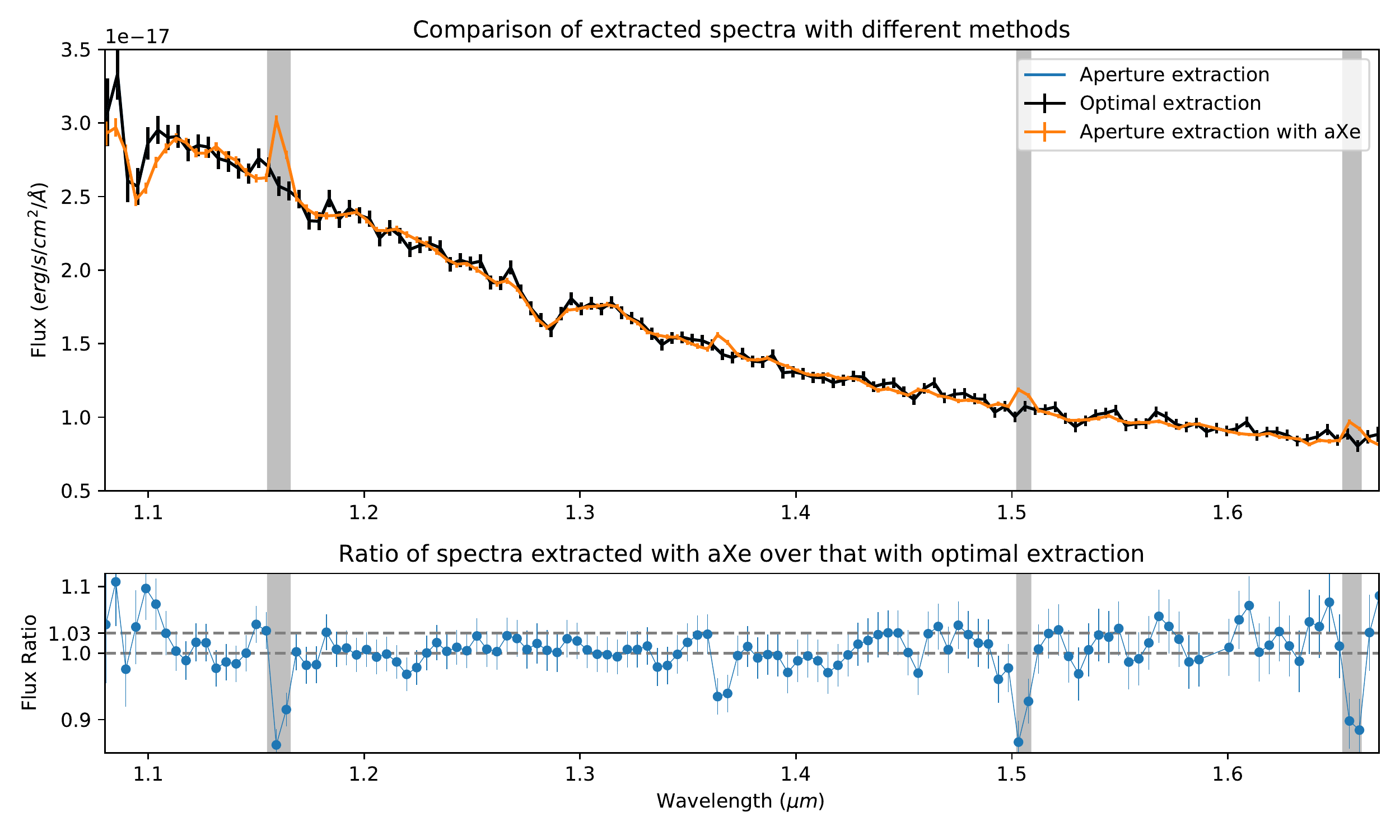}
    \caption{Comparison of median spectra with different spectral extraction methods illustrate that the three spectra extracted with different methods are consistent with each other within $\sim $3\% levels for wavelength range of 1.1--1.67 $\rm \mu m$. 
    The uniform and optimal extraction methods do not change the median spectra but do change the standard deviation of the light curve.
    The bottom panel shows the ratio of spectra extracted with the self-developed pipeline and with the standard aXe pipeline.
    visual inspection of the images confirm that the three grey shaded regions of the spectra extracted with aXe pipeline are affected by cosmic-ray or bad pixels.}
    \label{fig:axe}
\end{figure}
We conduct several tests to benchmark our extracted spectra. We confirm that the spectra extracted with optimal extraction method are consistent with apertures of 3, 4, and 5 pixels.
The median differences in the flux density between two spectra extracted with different apertures are less than 0.6\%, which is about five times lower than the averaged observed uncertainty of 3\%.
We then compare the spectra that are extracted with the new pipeline and that with the standard aXe pipeline that have been used in many studies \citep[e.g.,][]{buenzli2012,apai2013,yang2015,lew2016,manjavacas2017,zhou2018}.
We compare the median spectra in the second HST-orbit observation that is free from eclipse event because an unresolved eclipse event in the standard pipeline will skew the median values of spectra.
We notice that the hydrogen Paschen absorption line at 1.09\um is off by around 10\% between the spectra extracted with aXe and with our own pipeline.
We interpret that the deviation is likely caused by the imperfect wavelength calibration.
Overall, our comparison suggest that our pipeline produce similar reduced spectra as that by standard aXe pipeline.
\section{Conversion from counts-per-pixel to counts-per-wavelength}\label{sec:weighting}
As mentioned in Section \ref{sec:reduction}, we convert the count-per-pixel to counts-per-wavelength with a weighting function.
The weighting function comprises two components,
$A_1$ and $A_2$.
$A_1$ projects each pixel to the wavelength solution in cross-dispersion direction and calculates the fractional pixel area that falls in the bins of the solution.
$A_1 (i,j) = 1 $ means that the entire area of pixel $i$ falls into the wavelength bin $j$ in the cross-dispersion direction.
$A_2$ assigns weights to pixels that are within a specified aperture width from the spectral trace.
The value of $A_2(i)$ ranges from zero to unity, with being unity means that the pixel $i$ is at a distance less or equal than the aperture width from the spectral trace.
The weighting function $W(i,j)$ of pixel $i$ at wavelength $j$ thus equals to $W(i,j)$ = $A_1(i,j) \times A_2(i) $.
We verify that the calculation result of weighting function by examining the total weight of each pixel $i$ over all wavelength bins is equal to unity and the total weight of each wavelength bin $j$ over all the pixels is equal to two times of the aperture width.
With the weighting function, we sum the weighted electron count over all pixels as count per wavelength bin $C(j) = \sum_{i=1}^{N} W(i,j) \times C(i) $, where $C(i)$ is the count rate of pixel $i$ and $C(j)$ is the count rate of wavelength bin $j$.
\section{The priors and posterior distribution of MCMC sampling results}\label{sec:mcmc}
\begin{figure}[!h]
    \centering
    \includegraphics[width=1.0\textwidth]{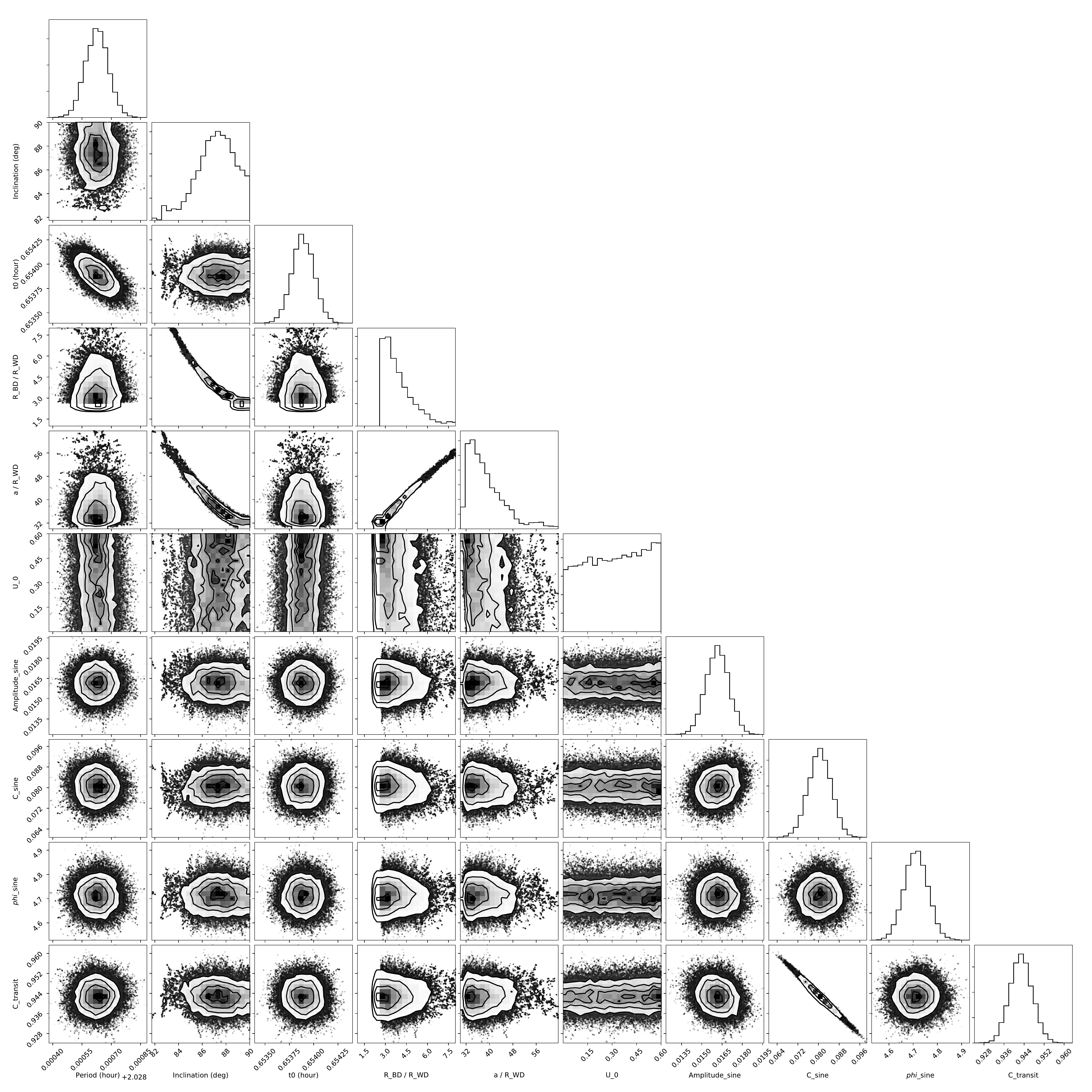}
    \caption{The posterior distribution of the light curve model parameters with MCMC method.}
    \label{fig:mcmc}
\end{figure}

Table \ref{table:priors} shows the uniform priors of the MCMC method used in the light curve modeling in Section \ref{sec:mideclipse}; Figure \ref{fig:mcmc} shows the posterior distribution of the MCMC modeling results.
\begin{table}[!h]
\raggedright
\begin{tabular}{ll}
\hline
\hline
Parameter & Prior range \\ \hline
Orbital period P (hour) & [2.0288*0.99, 2.0288*1.01]        \\ 
Radius ratio $r_{\text{BD}}/r_{\text{WD}}$  &  [1, $\frac{1.4
R_{\text{Jup}}}{0.012 R_{\odot}}$ = 14]          \\ 
Inclination $i$ (deg)      & [80, 90] \\ 
Mid-eclipse time $t_0$ (hour)& [0.63, 0.68]       \\ 
Semi-major-axis ratio a/$r_{\mathrm{WD}}$ & [30, 70]\\ 
Limb-darkening coefficient &  [0, 0.6]  \\
Flux constant $c_{\mathrm{WD}}$ & [0.8, 1]       \\ 
$A_{\mathrm{sine}}$    & [0, 0.1]        \\ 
$C_{\mathrm{sine}}$ & [0, 0.2]\\
$\phi_{\mathrm{sine}} $ & [$10^{-5}$, 2]$\pi$             \\ 
\hline \hline
\end{tabular}
\caption{The uniform priors of the light curve model parameters in the MCMC method. } \label{table:priors}
\end{table}

\bibliography{reference}{}



\end{document}